\documentclass[10pt,a4paper]{article}  % only supports two-column, letterpaper format
\usepackage{makeidx}
\usepackage{amssymb,amsfonts,amsmath,amstext}
\usepackage{epsfig}
\usepackage{textcomp}
\usepackage{fullpage}
\usepackage{watermark}
\usepackage[table]{xcolor}
\usepackage{multicol}
\usepackage{pstricks,pst-text,psfrag}
\usepackage{multirow}
\usepackage{cases}
\usepackage[sectionbib, round]{natbib}
\def\b1{{\mathbf 1}}

\newcommand{\bphi}{\boldsymbol\phi}

\def\bphi{{\mbox{\boldmath{$\phi$}}}}

\newcommand{\cC}{\ensuremath{\mathcal{C}}}

\def\cf{{cf.\ }}

\def\eg{{e.g.,\ }}
\def\ie{{i.e.,\ }}

\newcommand{\betab}{\begin{tabbing}}
\newcommand{\entab}{\end{tabbing}}
\newcommand{\beitem}{\begin{itemize}}
\newcommand{\enitem}{\end{itemize}}
\newcommand{\bea}{\begin{array}}
\newcommand{\ena}{\end{array}}
\newcommand{\beq}{\begin{equation}}
\newcommand{\enq}{\end{equation}}
\newcommand{\beqa}{\begin{eqnarray}}
\newcommand{\enqa}{\end{eqnarray}}
\newcommand{\beqan}{\begin{eqnarray*}}
\newcommand{\enqan}{\end{eqnarray*}}
\newcommand{\beenum}{\begin{enumerate}}
\newcommand{\enenum}{\end{enumerate}}
\newcommand{\DL}{\begin{dashlist}}
\newcommand{\DLE}{\end{dashlist}}

\newcommand{\bd}{{\ensuremath{\mathbf{d}}}}

\newcommand{\br}{{\ensuremath{\mathbf{r}}}}

\newcommand{\bx}{{\ensuremath{\mathbf{x}}}}

\newcommand{\bX}{{\ensuremath{\mathbf{X}}}}
\newcommand{\bY}{{\ensuremath{\mathbf{Y}}}}

\def\bpsi{{\mbox{\boldmath{$\psi$}}}}

\def\wrt{{w.r.t.\ }}

\newcommand{\Cramer}{Cram\'{e}r}

\makeatletter
\renewcommand*\env@matrix[1][c]{\hskip -\arraycolsep
  \let\@ifnextchar\new@ifnextchar
  \array{*\c@MaxMatrixCols #1}}
\makeatother

\newcommand{\ssim}{\raisebox{2pt}{$\scriptstyle \sim$}}

\newcommand{\theTitle}	       {\Large Space-based Aperture Array For \\ Ultra-Long Wavelength Radio Astronomy}

\title{\theTitle}
\author{
Raj Thilak Rajan$^{1, 2}
\thanks{
Correponding author: R.T.Rajan, rrajthilak aT gmail dOt com, \newline
$^{1}$ Netherlands Institute for Radio Astronomy (ASTRON), Dwingeloo, The Netherlands, \newline
$^{2}$ TU Delft, Delft, The Netherlands, \newline
$^{3}$ University of Twente, Enschede, The Netherlands, \newline
$^{4}$ Radboud University, Nijmegen, The Netherlands, \newline
$^{5}$ Airbus Defence $\&$ Space, Friedrichshafen, Germany}
$\and
Albert-Jan Boonstra$^{1}$\and
Mark Bentum$^{3,1}$ \and
Marc Klein-Wolt$^{4}$\and 
Frederik Belien$^{5, 2}$\and
Michel Arts$^{1}$ \and
Noah Saks$^{5}$ \and
Alle-Jan van der Veen$^{2}$ 
}
\begin{document}
\date{\today}
\maketitle

\begin{abstract}  The past decade has seen the rise of various radio astronomy arrays, particularly for low-frequency observations below $100$MHz. These developments have been primarily driven by interesting and fundamental scientific questions, such as studying the dark ages and epoch of re-ionization, by detecting the highly red-shifted $21$cm line emission. However, Earth-based radio astronomy below frequencies of $30$MHz is severely restricted due to man-made interference, ionospheric distortion and almost complete non-transparency of the ionosphere below $10$MHz. Therefore, this narrow spectral band remains possibly the last unexplored frequency range in radio astronomy. A straightforward solution to study the universe at these frequencies is to deploy a \emph{space-based antenna array} far away from Earths' ionosphere. Various studies in the past were principally limited by technology and computing resources, however current processing and communication trends indicate otherwise. Furthermore, successful missions which mapped the sky in this frequency regime, such as the lunar orbiter RAE-2, were restricted by very poor spatial resolution. Recently concluded studies, such as DARIS (Disturbuted Aperture Array for Radio Astronomy In Space) have shown the ready feasibility of a $9-$satellite constellation using off the shelf components. The aim of this article is to discuss the current trends and technologies towards the feasibility of a space-based aperture array for astronomical observations in the Ultra-Long Wavelength (ULW) regime of greater than $10$m \ie  below $30$MHz. We briefly present the achievable science cases, and discuss the system design for selected scenarios, such as extra-galactic surveys. An extensive discussion is presented on various sub-systems of the potential satellite array, such as radio astronomical antenna design, the on-board signal processing, communication architectures and joint space-time estimation of the satellite network. In light of a scalable array and to avert single point of failure, we propose both centralized and distributed solutions for the ULW space-based array. We highlight the benefits of various deployment locations and summarize the technological challenges for future space-based radio arrays.
\end{abstract}

%\keywords{Radio astronomy, Ultra-long wavelength, Interferometry, Feasibility study, System design}

%These arrays include the LOw Frequency Array (LOFAR) operating at $10-240$ MHz \citep{vanHaarlem2013}, MWA (Murchison Widefield Array) operating at $80-240$ MHz \citep{lonsdale2009}, LWA (Long Wavelength Array) operating at $10-88$MHz \citep{ellingson2009} and the Nan{\c{c}}ay decameter array operating at $20-70$MHz \citep{lecacheux2000} to name a few. 

%and the Nan{\c{c}}ay decameter array \citep{lecacheux2000}

\newpage
\section{Introduction} The success of Earth-based radio astronomy in the frequencies between $30$MHz and $3$GHz, is jointly credited to Earth's transparent ionosphere and the steady technological advancements during the past few decades. In recent times, radio astronomy has seen the advent of a large suit of radio telescopes, particularly towards the longer observational wavelengths \ie $\ge 3$m. These arrays include the Murchison Widefield Array (MWA) \citep{lonsdale2009}, LOw Frequency Array (LOFAR) \citep{vanHaarlem2013} and the Long Wavelength Array (LWA) \citep{ellingson2009} to name a few. However, Earth-based astronomical observations at these ultra-long wavelengths are severely restricted \citep{kaiser2000}. Firstly, due to ionospheric distortion, especially during the solar maximum period, when scintillation occurs and the celestial signals suffer from de-correlation among the elements of a ground based telescope array \citep{kassim1993}. Advanced calibration and mitigation techniques which are currently employed in LOFAR telescope array, can be used to remove these distortions, provided the time scales of disturbances is much longer than the time needed for calibration process \citep{wijnholds2010calibration}. Furthermore, at frequencies below $10$MHz the ionosphere is completely non-transparent which impede observations by ground-based instruments. In addition to ionospheric interference, man-made transmitter signals below $30$MHz also impede astronomical observations. This terrestrial interference was even observed as far as \ssim$400,000$km away from Earth by the RAE-2 lunar orbiter,  which was limited by very poor spatial resolution at these wavelengths, \eg $37^{\circ}$ at $9.18$MHz \citep{alexander1975}. Due to the above mentioned reasons, the very low frequency range of $0.3-30$ MHz remains one of the last unexplored frontier in astronomy. \emph{An unequivocal solution to observe the radio sky at ULW with the desired resolution and sensitivity is to deploy a dedicated satellite array in outer-space.} Such a space-based array must be deployed sufficiently far away from Earths' ionosphere, to avoid terrestrial-interference and offer stable conditions for calibration during scientific observations.

% The promise of ultra-long wavelength astronomy has lured many for over half a century now
\subsection{Science at ultra-long wavelengths} \label{sec:science} A space-based low frequency radio instrument would open up the virtually unexplored ULW domain and as such addresses a wealth of science cases that undoubtedly will lead to new exciting scientific discoveries, similar to the uncovering of other wavelength domains has revealed in the past. For most of these science cases such an array would add information to the existing radio, optical, infrared, sub-mm or high frequency X-ray or gamma-ray instrumentation, and thereby providing insight into the processes that take place at the lowest energies and largest physical scales. As explained in \citep{jester2009} and \citep{wolt2012} science topics include for example the study of the solar activity and space weather providing important clues on the effect of solar flares and bursts on the Earth to much larger distances from the solar surface. Investigating the magnetospheric emission from large planets such as Jupiter and Saturn \citep*[see][]{zarka2012} reveals information on the spin period of these planets. Other key science cases include the study of large-scale structures from galaxy clusters (and radio galaxies) and the detection of Jupiter-like flares and Crab-like pulses from (extra-)galactic sources. 

The greatest advance in science is expected in the study of the very early universe in a period referred to as the cosmological Dark Ages \citep{Rees99}. The only direct window to this period in the evolution of the early universe is provided by the 21-cm line from the neutral hydrogen caused by the spin-flip of the electron which, redshifted up to a factor 1000, is now visible in the low frequency radio regime between 1.4 - 140 MHz. The Dark Ages is the period between the epoch of recombination when the universe became transparent and the cosmic (microwave) background was emitted, and the epoch of reionisation (EoR) when the first stars started to reionise the neutral hydrogen. The hydrogen has played a major role in the Dark Ages and the EoR, being the only source of light in the absence of stars and through couplings with the cosmic background radiation and the surrounding neutral hydrogen it allows us to trace the distribution of matter in the early universe and observe the formation of the very first stars and structures in the universe \citep{CiardiFerrara05}. The global Dark Ages signal, essentially the redshifted 21-cm line absorption feature, is expected to peak around 30-40 MHz and is weak, $\sim$10$^{6}$ below the foreground signal. 

However, \citep{jester2009} show that with a single antenna placed at an ideal location on the moon (i.e. under low RFI and stable temperature and gain conditions) the global signal can be detected at a 5 $\sigma$ level for one year of integration. In order to trace the variations in the hydrogen at arc-minute or even arc second scale resolution a larger sensitivity and hence collecting area is required. For instance, \citep{jester2009} show that up to 10${5}$ individual antenna elements are required corresponding to 0.5 km$^{2}$ in order to reach $10'$ spatial resolution \citep*[see][]{loeb2004}. So, in addition to a low-RFI, low-temperature and stable gain location, a ULW radio interferometer that aims at detecting now only the global Dark Ages signal but also the arcmin variations in the mass distribution of the Dark Ages and the EoR must have a large collecting area (in the order of km$^{2}$). Ideal locations include the farside of the moon, an eternally dark crater on the lunar south or north pole, or space-based solutions such as for instance in a Sun-leading or trailing orbit, at the Sun-Earth L2 point or in Lunar orbit. 

%Several concepts have been suggested and investigated, from Lunar-based concepts such as the LRX instrument on the European Lunar Lander \citep{wolt2012}, the FarSide explorer \citep{mimoun2012farside}, the Dark Ages Radio Explorer (DARE), to space-based concepts such as DARIS and Dark Ages eXplorer (DEX). In this article we particularly focus on a constellation of less than $10$ satellites. However for arrays containing larger than $10$ satellites, scalable solutions have been offered in studies such as \eg OLFAR. The biggest advantage of a scalable concept is that one can do science from day one, detecting the global signal with a demonstrator mission with a single antenna, and then increase the sensitivity and hence the science by introducing more nodes later on.

%For all interferometric concepts, i.e. requiring multiple antenna elements, light-weight and scalable solutions have been offered. These include swarms of nano-satellites (e.g. OLFAR), inflatable balloon-like structures or solar-sail like structures (DEX) to thin foils that have antennas etched on them and that are rolled out on the surface of the moon (DARE). 

\subsection{Previous studies} The proposition for a space-based radio astronomy instrument is not novel \citep{weiler1988low, basart1997a, basart1997b, kaiser2000}. One of the first such proposal was by \citet{gorgolewski1965}, who discussed the benefits of a moon-based radio interferometer. In $1973$ the lunar orbiter RAE-2 was launched, which mapped the non-thermal galactic emission in the frequency range of $25$kHz to $13$MHz using a $37$m dipole antenna and achieving a resolution of $~37^{\circ}$ \citep{alexander1975}. Science at the Ultra-long wavelengths was revived from $1980$s with a particular focus on Lunar based arrays \citep{burke1990, burns1990observatories}. The Lunar surface on the far-side presents a large and stable platform for antennas and shields unwanted interference from Earth and the Sun \citep{woan1999very, kuiper2000lunar,takahashi2003concept,aminaei2014}, which motivated studies such as VLFA \citep{smith1990vlfa}, MERIT \citep{jones2007merit} and more recently DEX \citep{wolt2013dex}. Along similar lines, lunar orbiting single-satellite missions dedicated for radio astronomy such as LORAE \citep{burns1990lorae} and DARE \citep{burns2012dare} were also investigated to map bright sources and to facilitate relatively easier Earth-based down-link of science data. Furthermore, the pursuit of higher angular resolutions has led to Earth-orbiting single-satellite missions such as HALCA \citep{hirabayashi2000halca} and Radio Astron \citep{kardashev2013} which enable Earth-space very long baseline interferometry.

The concept of space based ULW array for radio astronomy however has seldom been explored adequately, which is our focus in this article. A notable study in this regard was the ALFA concept, which proposed an array of $3-16$ satellites in the radio quiet L2 points as a deployment location \citep{jones1998}. More recently, two ESA funded studies namely FIRST \citep{bergman2009} and DARIS \citep{boonstra2010daris} investigated passive-formation flying missions for space-based satellite arrays (see \tablename\ \ref{tb:studies}). The FIRST study proposed a constellation of $7$-satellites deployed at the second Earth-Moon Lagrange (L2) point, which allowed for a low-drift orbit and yet remained sufficiently far enough from Earth to avert interference. On the other hand, the DARIS mission primarily investigated the feasible ULW science cases and showed ready feasibility of $9$-satellites using existing off the shelf technologies. The benefits of both the studies are combined in the SURO concept, which proposes a mission with a larger observational frequency range at Sun-Earth L2. In all these studies, a dedicated centralized mothership managed the processing and communication. However, futuristic arrays such as OLFAR with $\ge 10$ satellites will operate cooperatively and employ a distributed architecture for both processing and communication.

%\citep{weiler1985astro} 	% AstroArray

\begin{table*}[!t]
\begin{center}
\resizebox{\textwidth}{!}{ \footnotesize
\begin{tabular}{|l|l|l|l|l|}
\hline
 & \textbf{FIRST} & \textbf{DARIS}  & \textbf{SURO-LC} &  \textbf{OLFAR} \\
 \hline
Timeline          									&  $2009-2010$& 	$2009-2010$             & $2011-2012$ &  $2010-2014$   \\
No. of satellites ($N$)             &  $6^{\dagger}+1^{\ddagger}$& $8^{\dagger}+1^{\ddagger}$  & $\ge 8^{\dagger}+1^{\ddagger}$&   $\ge 10$  \\
No. of polarizations ($N_{pol}$)    &  $3$  &$3$                      & $3$  &   $3$  \\
Obs. frequency ($\nu$)    					&  $0.3 - 50$ MHz& 	$0.3 - 10$ MHz       	& $0.5 - 60$ MHz 	&   $0.3- 30$ MHz  \\
Instantaneous BW ($\Delta \nu$)			&  $100$KHz     &  $1$ MHz           	& $1$MHz					&   $\ge 1$ MHz \\
Obs. wavelength ($\lambda$)					& $600- 6$ m	& $10^3- 30$ m         	& $600- 5$m 			&   $10^3- 10$m 	  \\
Longest baseline                    &  $30$ km &  $100$ km                 & $30$ km 				&   $100$ km    \\
Spatial resolution									&  $0.6′$ at $50$MHz &  $1'$ at $10$MHz  & $0.5'$ at $60$MHz &   $ 0.3'$ at $30$MHz \\
Array architecture        					&  Centralized &  Centralized              & Centralized 		&     Distributed   \\
Estimated Mass											& $\sim200$Kg$^{\dagger}$, $\sim10$Kg$^{\ddagger}$ 
																		& $\sim550$Kg$^{\dagger}$, $\sim100$Kg$^{\ddagger}$ 
																		& $\sim500$Kg$^{\dagger}$, $\sim10$Kg$^{\ddagger}$  
																		& $\le 5$Kg \\
Deployment location(s)      				& Sun-Earth L2 								&  Dynamic solar orbit, 		& Sun-Earth L2		&     Earth/Moon orbit \\
                                    & 														&  Moon far side,           &     & \\
                                    & 														&  Sun-Earth L2             &     & \\
\hline
\end{tabular}
}
\caption{\footnotesize \textbf{Recent Space-based aperture array studies:} An overview of system requirements for various space-based aperture array feasibility studies for ultra-long wavelength observations, namely DARIS \citep{saks2010,boonstra2010daris}, SURO-LC \citep{baan2012}  and OLFAR \citep{bentum2009olfar,rajan2011ac}, where $\dagger$ and $\ddagger$ denote  mothership and daughter node respectively.} 
\label{tb:studies}
\end{center}
\end{table*}
%Snapshot integration time           &  $1- 1000$ s,&  $1-1000$ s, & $1-1000$ s\\ %dependent on deployment location \\

\subsection{Overview} The purpose of this article is to discuss \emph{the current technological advances towards the feasibility of space-based array for radio astronomy at ultra-long wavelengths}. To this end, we elaborate on the system design for a space-based array in Section \ref{sec:system}. We address various subsystems of the potential satellite array in the Sections \ref{sec:science} - \ref{sec:syncLoc}, including the astronomy antenna design in Section \ref{sec:antennas}. While current technologies limits us to $\le 10$ nodes, we foresee next generation arrays will contain larger number of satellites and operate as a co-operative wireless network. Hence, a dominant theme of the article is to investigate the extension of the proposed centralized solutions to distributed scenarios, particularly for processing (Section \ref{sec:processing}), communication (Section \ref{sec:communication}) and joint space-time estimation of the satellites in the network (Section \ref{sec:syncLoc}). We summarize the article with a brief overview of the potential deployment locations (Section \ref{sec:deploymentLocations}) and the fundamental challenges ahead for a space-based ULW array (Section \ref{sec:challenges}).

\section{Ultra-long wavelength interferometry} \label{sec:system}
\subsection{Aperture synthesis} \label{sec:system:apertureSynthesis} 
\begin{figure*}[t!]%
\centering

\psfrag{uT0}[cb]{\small (a) uvw: Snapshot}
\psfrag{uT1}[cb]{\small (c) uvw: Bandwidth synthesis}
\psfrag{uT2}[cb]{\small (e) uvw: Complete orbit}
\psfrag{uX0}[cc]{\small u}
\psfrag{uY0}[lt]{\small v}
\psfrag{uZ0}[cc]{\small w}
\psfrag{uX1}[cc]{\small u}
\psfrag{uY1}[cc]{\small v}
\psfrag{uZ1}[cc]{\small w}
\psfrag{uX2}[cc]{\small u}
\psfrag{uY2}[cc]{\small v}
\psfrag{uZ2}[cc]{\small w}

\psfrag{bT0}[cb]{\small (b) PSF: Snapshot}
\psfrag{bT1}[cb]{\small (d) PSF: Bandwidth synthesis}
\psfrag{bT2}[cb]{\small (e) PSF: Complete orbit}
\psfrag{bX0}[tc]{\small $\theta_u$ [arcmin]}
\psfrag{bY0}[bc]{\small $\theta_v$ [arcmin]}
\psfrag{bX1}[tc]{\small $\theta_u$ [arcmin]}
\psfrag{bY1}[bc]{\small $\theta_v$ [arcmin]}
\psfrag{bX2}[tc]{\small $\theta_u$ [arcmin]}
\psfrag{bY2}[bc]{\small $\theta_v$ [arcmin]}

\parbox{2in}{\includegraphics[scale=0.23]{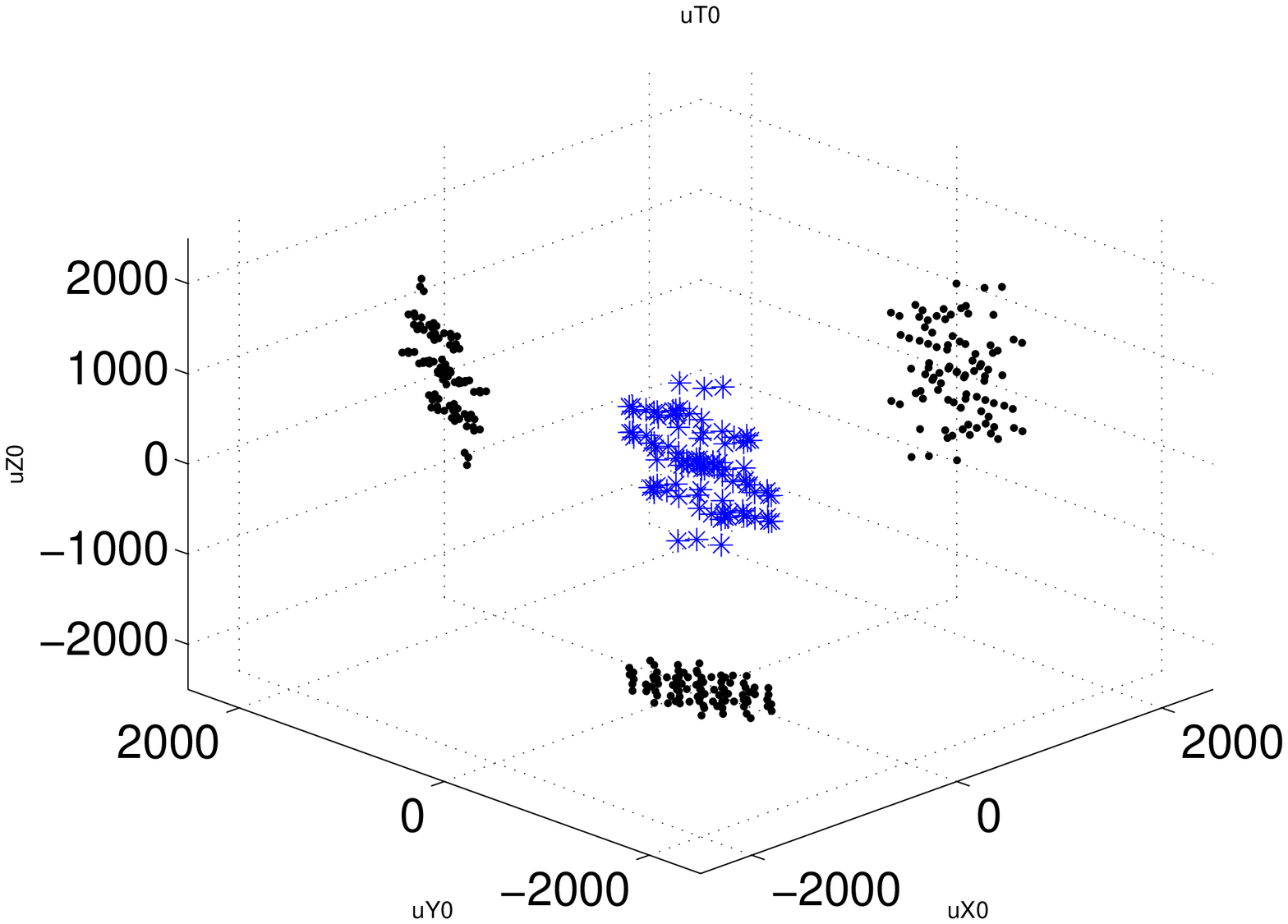}} \hspace{-5mm}
\parbox{2in}{\includegraphics[scale=0.23]{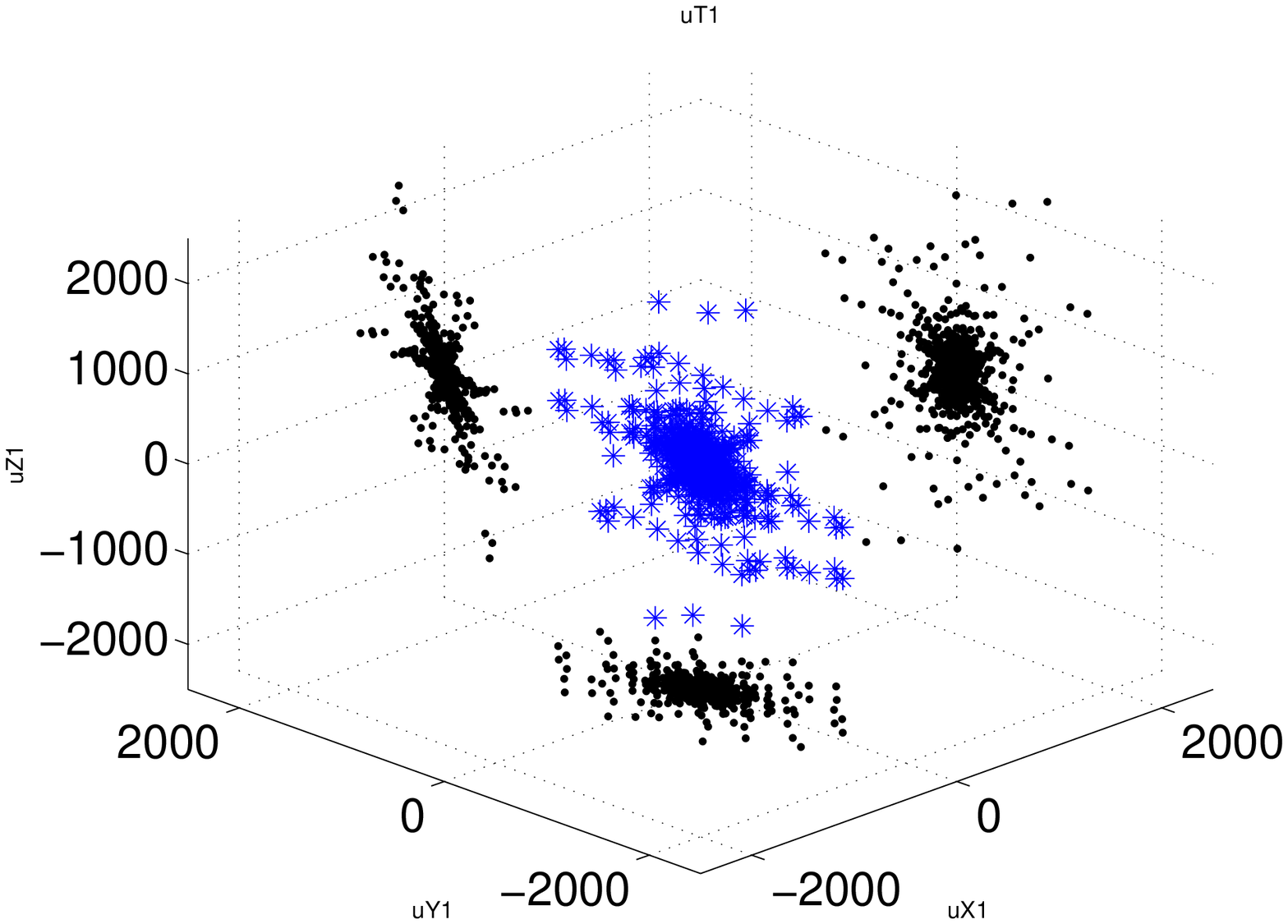}} \hspace{-5mm}
\parbox{2in}{\includegraphics[scale=0.23]{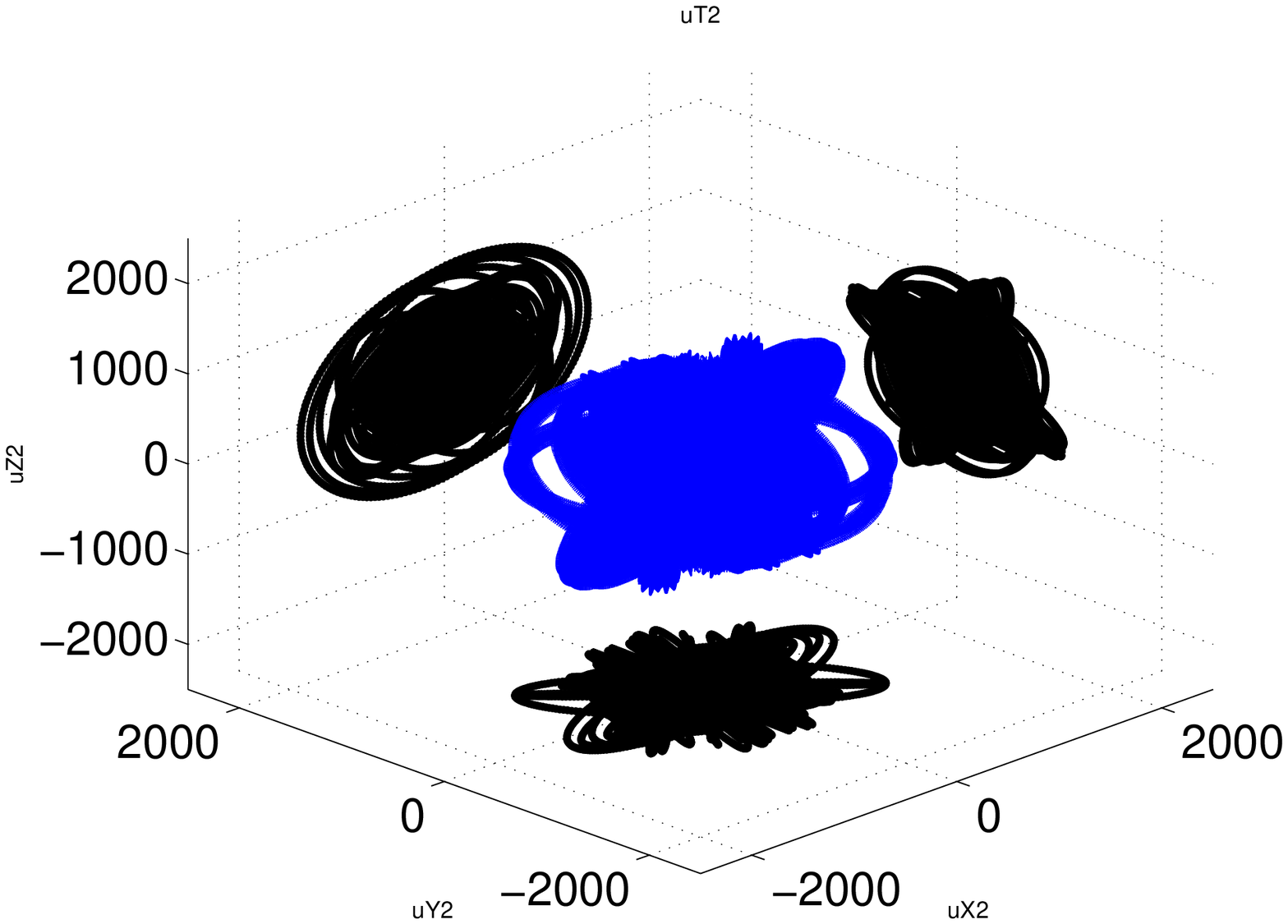}} \\ \vspace{7.5mm}
\parbox{2in}{\includegraphics[scale=0.23]{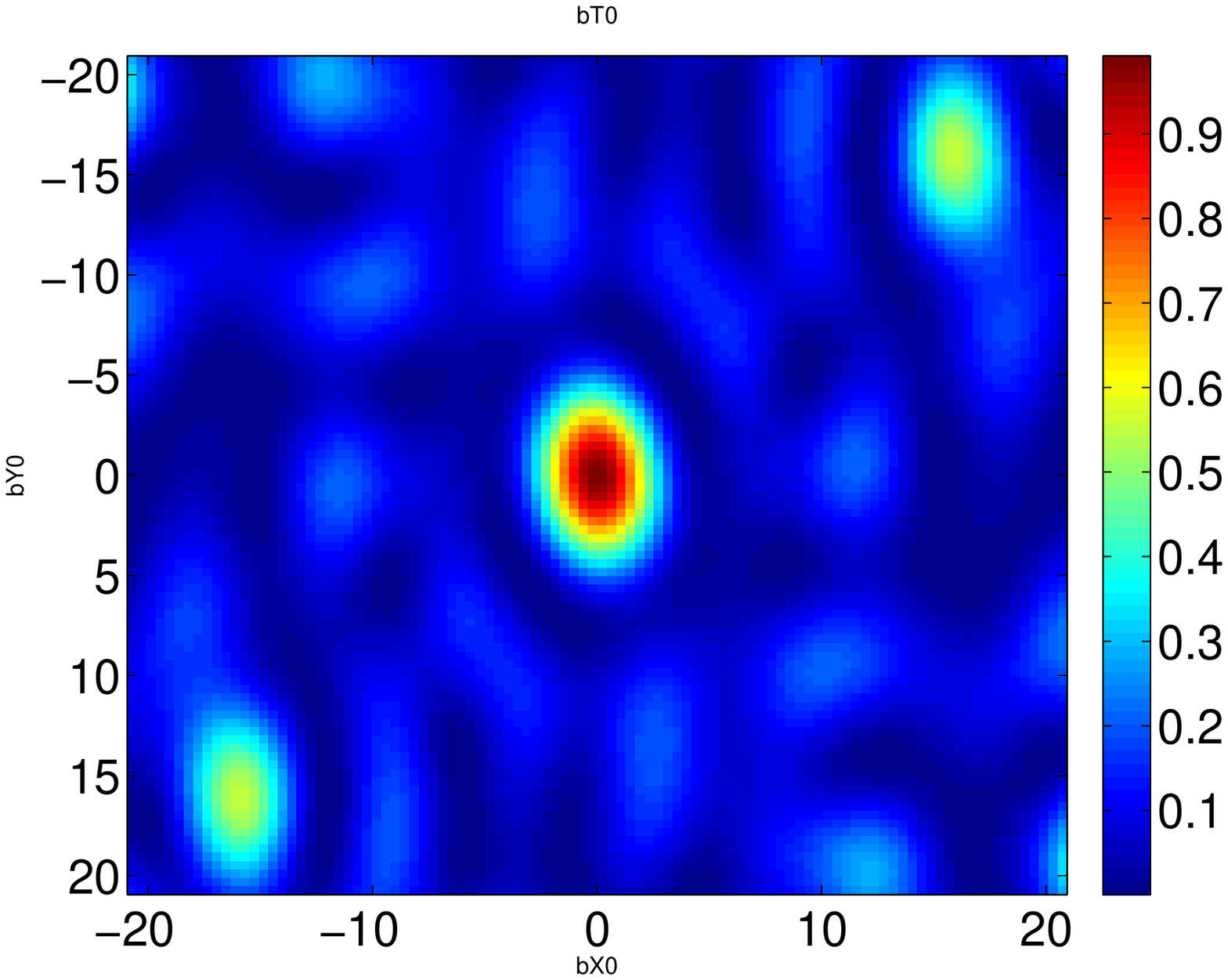}} \hspace{-5mm}
\parbox{2in}{\includegraphics[scale=0.23]{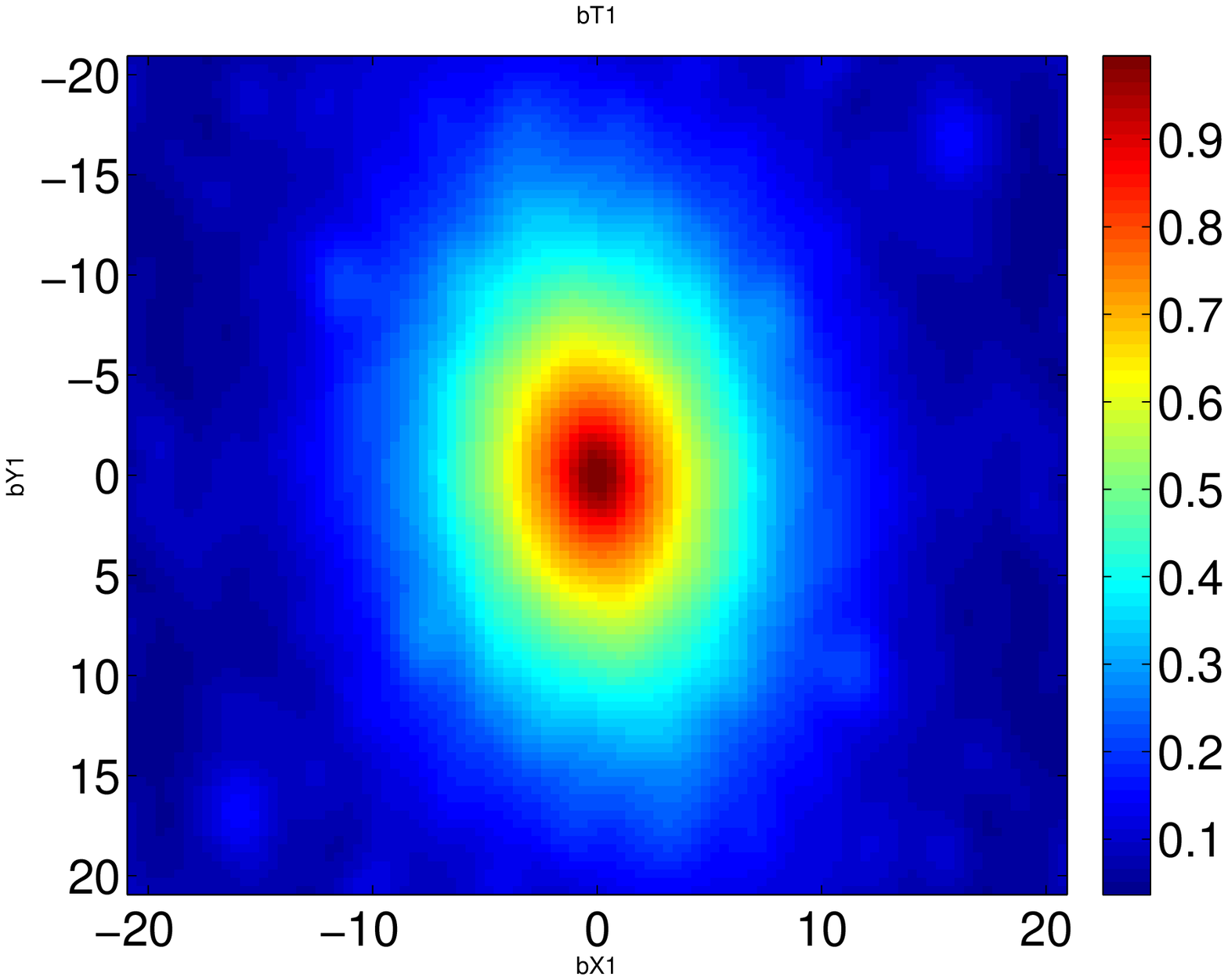}} \hspace{-5mm}
\parbox{2in}{\includegraphics[scale=0.23]{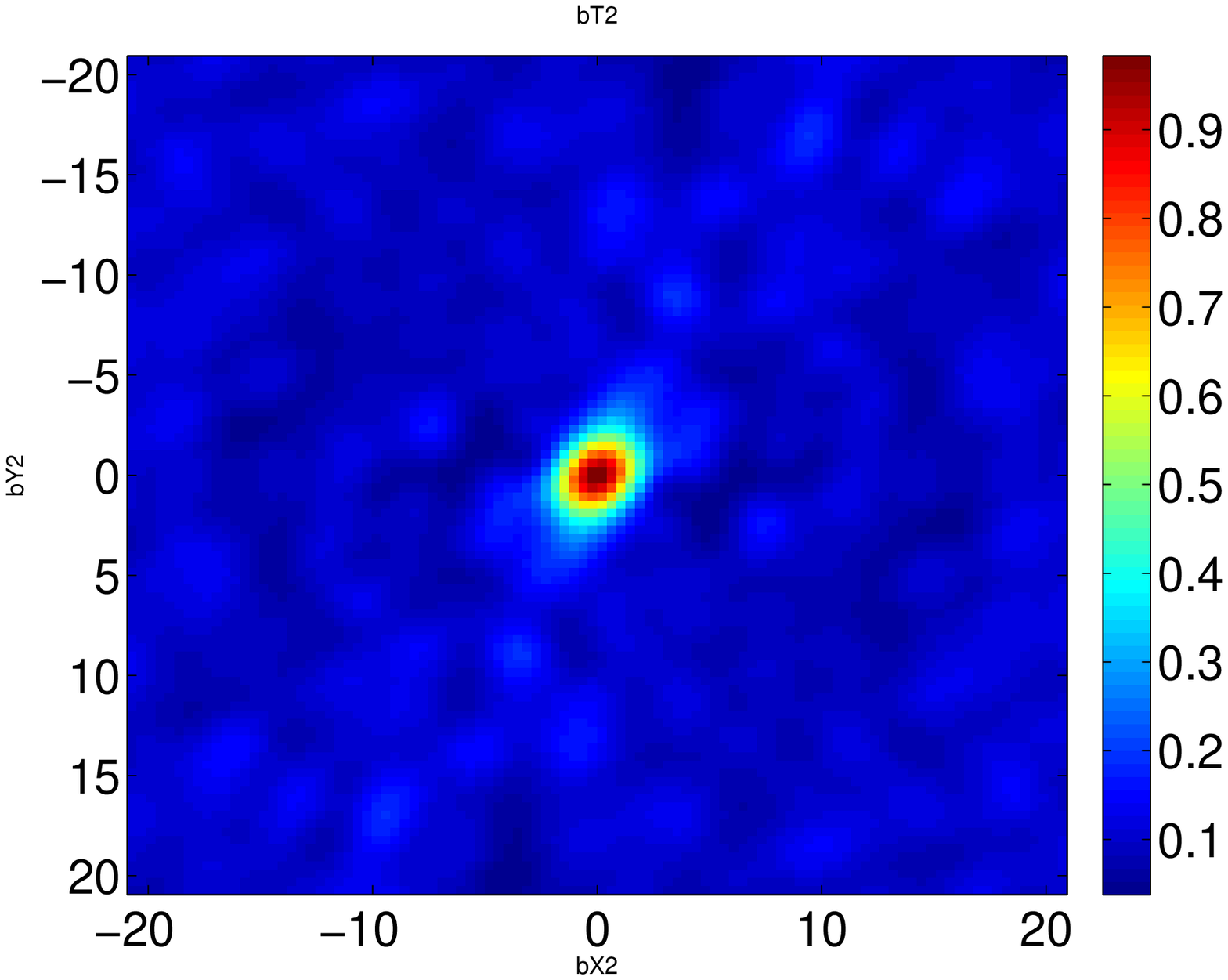}}

\vspace{0.5\baselineskip}
\caption{\footnotesize \textbf{\emph{Baseline and Sky map simulations: }} Aperture filling of a $9$-satellite ULW array for an Earth leading orbit around the Sun, to illustrate the effect of the sampling space on the normalized Point Spread Function (PSF). The uvw coverage for the $3$-D array of satellites at at $\nu=10$MHz (a) for a single snapshot $N_t=1$, (b) single snapshot using $10$ frequency bins uniformly distributed in the range $1$ - $10$ MHz and (c) for an entire orbit around the sun at $10$MHz with single observation each day \ie $365$ snapshots.}

%\caption{\footnotesize uvw + PSF}
\label{fig:uvwSimulation}%
\end{figure*}

Radio astronomy imaging is achieved by aperture synthesis, where in the cosmic signals received at a large number of time-varying antenna positions, are coherently combined to produce high quality sky maps. For a $N-$antenna array, each antenna pair forms a \emph{baseline} of an aperture synthesis interferometer, contributing $\bar{N} \triangleq\ 0.5N(N-1)$ unique sampling points at a given time instant. Let $\bx_i(t)$ and $\bx_j(t)$ be two arbitrary antenna position vectors at time $t $ forming a baseline, then the corresponding uvw point is defined as \begin{equation} \label{eq:uvw}  [u_{ij}(t), v_{ij}(t), w_{ij}(t)] \triangleq\ (\bx_i(t)-\bx_j(t))/\lambda, 
\end{equation} where $\lambda$ is the observed wavelength. \figurename\ \ref{fig:uvwSimulation}(a) shows the uvw for a $N=9$ satellite cluster which is arbitrarily deployed with a maximum distance separation of $d=50$km and an observational frequency of $10$MHz. The effective synthesized aperture is then obtained by projecting the uvw points onto a 2-D plane which is orthogonal to the source direction. As an illustration, \figurename\ \ref{fig:uvwSimulation}(a) shows $3$ such projections (in black) for sources orthogonal to the $uv$, $uw$ and $wv$ planes. The minimum distance between the satellites is only constrained by practical safety requirements and the maximum distance between the satellites defines the resolution of the interferometric array as \begin{equation}\label{eq:arrayResolution} 
\theta= \lambda/d. \end{equation}

The Van Cittert Zernike theorem relates the spatial correlation of these antenna pairs directly to the source brightness distribution by a Fourier transform \citep{thompson2008}. Hence for radio imaging, each antenna pair output is cross-correlated to measure the coherence function which is subsequently converted to a sky map, conventionally by an inverse Fourier transform. \figurename\ \ref{fig:uvwSimulation}(b) shows the normalized Point Spread Function (PSF) corresponding to the aperture coverage in \figurename\ \ref{fig:uvwSimulation}(a),  for a single point source along the $w$ direction. The more densely sampled the projected aperture plane, the lower the spatial side-lobes of the sky image. The filling factor of the synthesized aperture can be increased by either using bandwidth synthesis or by populating sufficient baselines. In bandwidth synthesis different frequency channels can be used to scale $\lambda$. As shown in \figurename \ref{fig:uvwSimulation}(c) and \figurename \ref{fig:uvwSimulation}(d), using only $10$ frequency bins uniformly distributed across $1-10$MHz, the aperture filling and the PSF is significantly improved as compared to \figurename \ref{fig:uvwSimulation}(b). A first-order simulation of an array of $N=9$ satellites in Earth-leading orbit around the sun yields \figurename \ref{fig:uvwSimulation}(e) and the corresponding PSF in \figurename \ref{fig:uvwSimulation}(f), where one snapshot each day is assumed at a single observation frequency of $10$MHz. The number of uvw points are directly related to the unique number of baselines and the observational frequency. To achieve the confusion limit and resolve the sources individually, the total number of unique uvw points over the observational time period must be larger than the total number of detected sources. 

% The total number of uvw points $N_{uvw}$ are directly related to the unique baselines and the observational frequency. However, under the assumption that the largest contribution is from the largest frequency \ie the shortest wavelength , a lower limit on the total number of uvw points can be estimated. More specifically, we have \begin{equation} \label{eq:numberOfuvw}
%N_{uvw} = 0.5N(N-1)t_{obs}\nu^{-1}\dot{r}\sin(50^o)
%\end{equation} where $t_{obs}$ is the observational time, $\dot{r}$ is the relative range rate between the satellites and we assume an average observation angle of $50^o$. \emph{For any observation, the total number of unique baselines must be larger than the total number of detected sources, in order to resolve the sources individually.} 

\subsection{Ultra-long wavelength sky} \label{sec:system:sky} The dominant foreground in the low frequency radio sky is the galactic synchrotron radiation, which is caused due to synchrotron emission from electrons moving in the Galactic magnetic field. This emission causes the brightness temperature to rise from \ssim $10^4$K at $30$MHz, to as high as \ssim $10^7$K around $2$MHz \citep{oberoi2005}. At frequencies below $2$MHz, the Galactic plane is nearly completely opaque and the extra-galactic sources cannot be observed. More explicitly, for frequencies above $2$MHz, the sky temperature can be approximated as \citep{jester2009} \begin{equation} \label{eq:Tsys} 
T_{sky}= 16.3 \times 10^{6} \text{K} \Big( \dfrac{\nu}{2MHz} \Big)^{-2.53} \quad \text{at}\ \nu > 	2 \text{MHz},
\end{equation} where $\nu$ is the observation frequency. For Earth-based observations at higher frequencies ($> 100$MHz), the overall system noise temperature $T_{sys}$ plaguing the cosmic signal is typically dominated by the noise from receiver electronics $T_{rec}$. However, at lower frequencies ($\le 30$ MHz), the intense galactic background implies that $T_{sky}$ will be at least an order magnitude larger than $T_{rec}$, and hence the overall noise temperature $T_{sky} \gg T_{sys}$. The immediate effect of this extremely high sky noise is the poor sensitivity of the interferometric array.  The $1$-$\sigma$ RMS sensitivity for an antenna array of $N$ nodes is \citep{cohen2004} \begin{equation}  \label{eq:arraySensitivity}
S_{\sigma}=	\frac{235.6\ T_{sys}}{\lambda^2\sqrt{N(N-1) (t_{obs}/1 \text{hour})(\Delta\nu/ 1\text{MHz})}} \text{mJy/beam}, 
\end{equation} where $\Delta \nu$ is the bandwidth, $t_{obs}$ is the observation time period over which the signal is integrated and the total number of estimated sources above this sensitivity is given by \begin{equation} \label{eq:numberOfSources}
N_{>}(S)= 1800 \big( S/10 \text{mJy} \big)^{-0.3} \big( \nu/ 10 \text{MHz} \big)^{-0.7}.
\end{equation} Furthermore, the scattering in the interplanetary media (IPM) and interstellar media (ISM) also hinder observational frequencies less than $30$MHz, which limit the maximum baseline between the satellites to \begin{equation} 
\label{eq:maxBaselineISM} d_{ISM}= 47 \text{km} \times (\nu/1\text{MHz})^{1.2} \qquad\ \text{and} \qquad\  d_{IPM}\approx 10 \text{km} \times (\nu/1\text{MHz}). \end{equation}

Finally, the lower limit of the achievable noise is not the RMS sensitivity of the array, but the confusion limit. The presence of unresolved sources with individual flux densities below the detection limit leads to a constant noise floor, that is reached after a certain observation time $t_{obs}$ \citep{jester2009}. For extra-galactic observations, under certain nominal assumptions, this anticipated confusion limit due to background sources is \begin{equation} 
\label{eq:confusionLimit}
S_{conf}(\theta, \nu)= 16\text{mJy} \times (\theta_{conf}/1')^{1.54}(\nu/74\text{MHz})^{-0.7},
\end{equation} where $\theta$ is the effective resolution for which the flux is below the confusion limit. The confusion limit is the lower limit to the achievable noise floor and thus is an upper limit to the useful collective area of the array. In other words, adding more antennas only decreases the time in which the confusion limit is reached, but not the overall array sensitivity (\ref{eq:arraySensitivity}). The time necessary for an array to reach achieve this confusion limited sensitivity is given by the ``survey equation'' \begin{equation} \label{eq:confusionTime}
t_{survey}= 3.3 \big(N/100)^{-2}
								\big(10\nu/\Delta \nu\big)
								\big(\nu/1\text{MHz}\big)^{-0.66}
								\big(\theta/1'\big)^{-3.08}.
\end{equation} Using these elementary and yet fundamental equations, a preliminary design for a space-based array can be proposed for desired science cases. For a more detailed study, refer to \citep{jester2009}.

% The science cases for an ULW array span from study of global epoch of re-ionization, global dark-ages signal, galactic and extra-galactic surveys, study of transients such as Solar and Planetary bursts, and even Ultra-high energy cosmic rays. 

\subsection{System definition} \label{sec:systemDefinition} The science cases for an ULW array broadly span cosmology, galactic surveys, transients from solar and planetary bursts and even the study of Ultra-High Energy particles. Although a single satellite mission would suffice to detect the global dark-ages signal, over $10^4$ antennas are required to investigate the radio emission from Extrasolar planets \citep*[see][ \tablename\ 1]{jester2009}. For a \emph{first space-based ULW array} however, with possibly only a few satellite nodes, extra-galactic surveys and study of transients are among the best suited science cases \citep{boonstra2011daris}, which we present as case studies. The proposed space-based array design can be readily extended to cater to other science cases \eg detection of global dark-ages signal.

\begin{table}[!t]
\resizebox{\textwidth}{!}{ \footnotesize
\begin{tabular}{|l|c|c|c|l|l|l|l|l|l|l|}
\hline
\multicolumn{1}{|c|}{\textbf{Parameter}} & {\textbf{Notation}} & \textbf{Units} & \textbf{Equation} & \multicolumn{6}{c|}{\textbf{Extra-galactic survey}} \\ \hline
Sensitivity &$S_{\sigma}$ & Jy & Input & 6.5E-02 & 6.5E-02 & 6.5E-02 &  6.5E-02 & 6.5E-02 & 6.5E-02 \\ \hline
Baseline &$d$& km & Input & 100 & 100 & 100 &   100 & 100 & 100 \\ \hline
Obs. Time &$t_{obs}$ & hours & Input & 24 & 720 & 8760 &   720 & 8760 & 8760 \\ \hline
Obs. frequency &$\nu$& MHz & Input & 10 & 10 & 10 &   10 & 10 & 30 \\ \hline
%Aeff &  & Derived & 225 & 225 & 225 &   & 225 & 225 & 25 \\ \hline
%Bandwidth \% &  & Input & 10 & 10 & 10 &   30 & 30 & 10 \\ \hline
Bandwidth &$\delta \nu$ & MHz & Input & 1 & 1 & 1 &   3 & 3 & 3 \\ \hline
Resolution &$\theta$ & arcmin &  (\ref{eq:arrayResolution}) & 1.03 & 1.03 & 1.03  & 1.03 & 1.03 & 0.34 \\ \hline
System temperature$^{\dagger}$ &$T_{sys}$ & K & (\ref{eq:Tsys}) & 2.8E+05 & 2.8E+05 & 2.8E+05 &  2.8E+05 & 2.8E+05 & 1.7E+04 \\ \hline
No. of Antennas &$N$ &  &  (\ref{eq:arraySensitivity}) & 229 & 42 & 12 & 25 & 7 & 4 \\ \hline
ISM Max. Baseline &$d_{ISM}$ & km &  (\ref{eq:maxBaselineISM}) & 7.4E+02 & 7.4E+05 & 7.4E+05 & 7.4E+05 & 7.4E+05 & 2.8E+06 \\ \hline
IPM Max. Baseline &$d_{IPM}$ & km & (\ref{eq:maxBaselineISM}) & 100.0 & 100.0 & 100.0  & 100.0 & 100.0 & 300.0 \\ \hline
Confusion limit &$S_{conf}$ & Jy & (\ref{eq:confusionLimit}) & 0.07 & 0.07 & 0.07  & 0.07 & 0.07 & 0.01 \\ \hline
Resolution (Conf. lim.) & & arcmin & (\ref{eq:arraySensitivity}), (\ref{eq:confusionLimit}) & 1.00 & 1.00 & 1.00  & 1.00 & 1.00 & 1.65 \\ \hline
Max. Baseline (Conf. lim.)& & km & (\ref{eq:arrayResolution}), (\ref{eq:confusionLimit}) & 103.09 & 103.09 & 103.09 & 103.09 & 103.09 & 20.86 \\ \hline
Time to Conf. lim. & & hours & (\ref{eq:confusionTime}),(\ref{eq:confusionLimit}) & 0.14 & 4.05 & 46.39 & 3.98 & 43.64 & 38.94 \\ \hline
%Sources detected&& (\ref{eq:numberOfSources}) & 6.5E+06&	6.5E+06&	6.5E+06&	6.5E+06&	6.5E+06&	3.0E+06 \\ \hline
%Number of uvw points & & (\ref{eq:numberOfuvw}) & 5.8E+07&	5.8E+07&	5.8E+07&	1.9E+07&	1.9E+07&	1.8E+07 \\ \hline

%No. of Antennas &  &  (\ref{eq:maxAntennaBaseline}) & 4.7E+07 & 4.7E+07 & 4.7E+07 & 4.7E+07 & 4.7E+07 & 4.7E+07 & 1.7E+07 \\ \hline
%\begin{tabular}[c]{@{}l@{}}Source\\   detection 1-sigma (All sky)\end{tabular} &  &  & 6.5E+06 & 6.5E+06 & 6.5E+06 & 6.5E+06 & 6.5E+06 & 6.5E+06 & 3.0E+06 \\ \hline
%\begin{tabular}[c]{@{}l@{}}Number of uv\\   points\end{tabular} &  &  & 8.0E+05 & 8.0E+05 & 8.0E+05 & 8.0E+05 & 8.0E+05 & 8.0E+05 & 3.7E+05 \\ \hline
\end{tabular}
}
\caption{\textbf{\emph{System requirements for Extra-galactic surveys:}} \footnotesize  The desired system parameters to achieve the desired resolution of $1'$ and sensitivity of $65$mJy for Extra-galactic surveys, for varying observation time, observation frequency and instantaneous bandwidth. ($\:^\dagger$ Sky noise dominated)}
\label{tb:systemDesign}
\end{table}

The expected signal strength for the extra-galactic surveys are in the order of $65$mJy with a desired spatial resolution of $\ssim 1'$. In \tablename\ \ref{tb:systemDesign}, we present different scenarios to investigate the effects of varying observational frequencies, bandwidth and observation time, on the number of antennas to achieve $65$mJy. It is evident that increasing the observation time ($1$ day, $1$ month, $1$ year) steadily reduces the required number of antennas. Secondly, the increasing the bandwidth ($1$MHz to $3$ MHz) is also an alternative to achieve the desired resolution for a small array. However, the increase in bandwidth has little effect on the confusion limit. We note that the confusion limit is a bottleneck for shorter integration times and lower observing frequencies. The maximum baseline is in general confusion limited for frequencies $\ge 10$MHz, however at $< 10$MHz, the ISM and IPM scattering limits the maximum baseline and subsequently the resolution. At $10$MHz, we require at least one year of observation time with more than $7$ antennas for an observational bandwidth of $3$MHz to achieve the $65$mJy sensitivity. However, in the last column of the \tablename\ \ref{tb:systemDesign}, we see that at $30$MHz, only $4$ antennas sufficient. Such a configuration is estimated to detect over a million sources using (\ref{eq:numberOfSources}). 

A similar investigation was conducted for Jupiter-like flares and Giant crab-like pulses, which are bright events with order of MJy and kJy respectively, with typical time scales of milliseconds. The desired resolution for these transient radio systems are $ \ssim 1'$ at $\le 30$MHz. Since these events are extremely bright, even a single antenna with a nominal bandwidth of $10\%$ of observational frequency would meet the desired requirements. However, unlike the extragalactic surveys the observations are not confusion limited but possibly by the number of baselines for short integration timescales of milliseconds. The number of unique uvw points will depend inadvertently on the deployment location and the relative range rate of the antennas. However, in general this limitation can be overcome by increasing the integration times in both these cases by over a minute.

In general, higher bandwidth, higher observing frequencies and longer integration times require less antennas to reach the same sensitivity level. In this article we choose the DARIS mission specifications as a reference to illustrate various sub-systems. To this end, we particularly focus on an array of $N=9$ satellites, with a maximum satellite separation of $100$km and capable of observing the skies at $0.1-10$MHz. This particular setup meets the requirements for the extra-galactic survey and transient radio system science cases. However, all the proposed techniques and technologies can be readily extended to a larger array, for observation frequencies up to $30$MHz.

%%-----------------------------------------------------------------------------------------------------------------------%

\begin{figure}[t!]  \centering
    \includegraphics[scale=0.22]{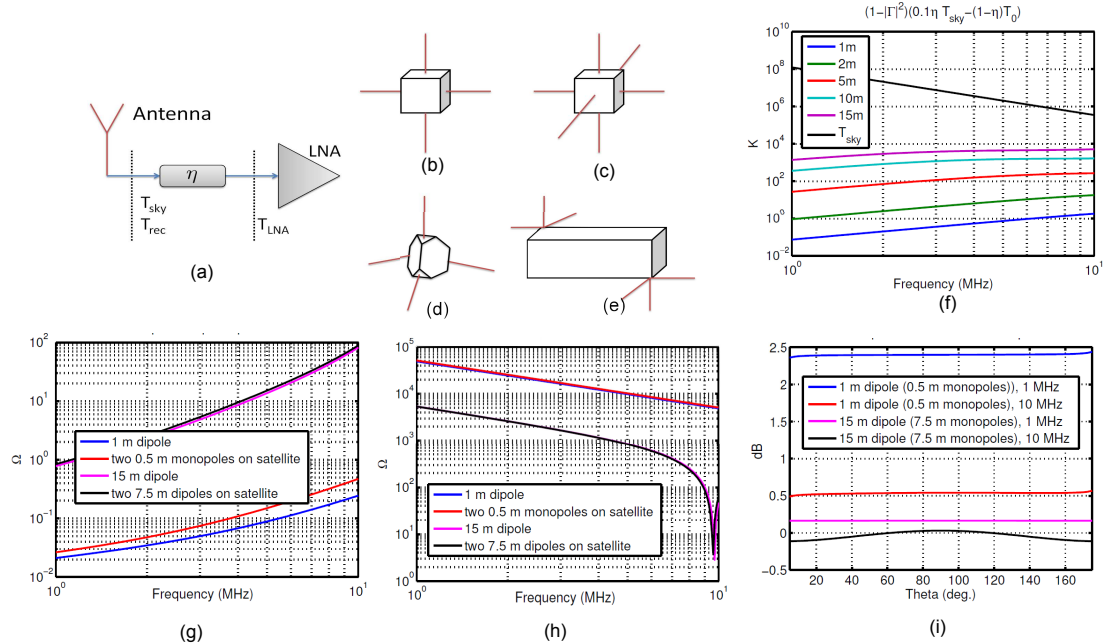}
    \caption{ \textbf{\emph{Space-based antenna model, configurations and simulations:}} \footnotesize (a) System model for an LNA connected to the Antenna (b) Configuration of a dipole antenna on a cubesat (c) Configuration of a tripole antenna in a cubesat (d) four monopoles (e) A dipole antenna placed on the opposite ends of a 3-cubesat (f) Required $T_{LNA}$ as a function of frequency for non-matched dipoles of varying lengths. (g) Real part of input impedance and (h) Absolute value of imaginary part of input impedance for two monopoles placed on a satellite compared to a dipole (i) Ratio of gain of two monopoles placed on a satellite and a dipole.}
\label{fig:antennas}
\end{figure}

\section{Radio astronomy antenna design} \label{sec:antennas} A critical component for the space-based array is the design of the observational antenna. The system model of the observation antenna connected to a LNA is shown in Figure \ref{fig:antennas}(a), wherein the antenna is modeled as an \emph{ideal} lossless antenna, followed by an attenuator representative of the antenna losses. For the observation frequencies of $0.3-30$MHz, this front end must be sky noise limited \ie $T_{rec} < 0.1T_{sky}$, where $T_{rec}$ is the receiver noise and $T_{sky}$ is the sky noise temperatures which are defined at the interface between the lossless antenna and attenuator. The LNA noise temperature $T_{LNA}$ defined at the input of the LNA is equal to $(1-\eta)T_0$, where $\eta$ is the radiation efficiency and $T_0$ is the physical temperature of the antenna (chosen as 290K). Without loss of generality, we assume that the LNA noise is dominant over the noise contribution of subsequent electronics of the receiver. Under these assumptions, the prerequisite on the LNA noise temperature is derived as \begin{equation} \label{eq:T_LNA} 
T_{LNA} < (1-|\Gamma|^2)(0.1\eta T_{sky}-(1-\eta)T_0 ), \end{equation} where $\Gamma$ is the reflection coefficient of the antenna \citep{arts2010}. A straightforward candidate for the observation antenna is a dipole (\eg \figurename \ref{fig:antennas}(b)), which can be realized by rolling out metallic strips from the satellite \citep{manning2000}. The observational wavelengths are much larger compared to the dimensions of the satellites and hence due to practical limitations, the realized dipole will be short compared to the wavelength. For instance, a classic half-wave dipole for $10$MHz and $30$MHz observation frequencies yields a dipole length of $15$m and $10$m respectively. For lower frequencies, a similar dipole lengths begets a short-dipole with a directional pattern similar but less directional as compared to the half-wave dipole. Consequentially, the radiation resistance of the small antenna would be low and the thermal noise will significantly dominate the total antenna noise. 

To maximize the received power at the antenna, matched dipoles can be used, in which case $\Gamma \approx\ 0$. However, the combination of antenna and matching network becomes highly resonant with a high quality factor, significantly limiting the achievable bandwidth of the system, in particular for shorter dipoles. Hence, we propose the use of a non-matched dipole \citep{arts2010}. Using (\ref{eq:T_LNA}), we have the \figurename \ref{fig:antennas}(f), which shows the $T_{LNA}$ for varying lengths of non-matched dipole lengths. We use a LNA with input impedance of $2$K$\Omega$, which is a realistic value for an LNA using a bipolar transistor. While the $T_{LNA}$ is above $100$K for $10$m and $15$m antennas, the required noise temperature for shorter dipole lengths are significantly lower, especially for lower frequencies.

%All antenna simulations are done with CAESAR (Computationally Advanced and Efficient Simulator for ARrays) \citep{maaskant2006caesar}. The dipoles are modeled as strips made from steel (conductivity $10^6$ S/m) with a thickness of $0.1$mm and a width of $1$cm.

In practice, a dipole is implemented on a satellite using two monopoles. To verify the validity of the proposed model, impedances of two monopoles on a satellite body is compared to that of the dipole. The satellite body under simulation is modeled as a cube of $40\times40\times40$cm , with perfect conducting surfaces. Furthermore, since the impedance of the monopole above a perfect ground plane is half the impedance of a dipole, the impedance of the monopole is multiplied by a factor $2$ in the simulations for a fair comparison. \figurename\ \ref{fig:antennas}(g) and \figurename\ \ref{fig:antennas}(h) show the real component of the impedance and absolute value of the imaginary component of the impedance respectively, for $1$m dipole versus $0.5$m monopoles and $15$m dipole versus $7.5$m monopoles respectively. As seen in these figures, there is negligible difference between the dipole and the two monopole configuration. A step further, we compare the ratio of gains between a pair of monopoles against the $1$ meter and $15$ meter dipoles at $1$ and $10$MHz. The investigated antenna lengths the ratio of gain are almost flat across for varying angles as seen in \figurename\ \ref{fig:antennas}(i), which indicates the element pattern does not change if a configuration of two monopoles is used instead of a dipole. 
% The sensitivity of two orthogonal dipoles is low in some directions for some polarizations

Two orthogonal dipoles (\figurename \ref{fig:antennas}(b)) are in principle sufficient to get all the polarization information of the cosmic signal, however a tripole (\ie three dipoles, see \figurename \ref{fig:antennas}(c)) can be used to obtain information of all $3$ components. The third dipole improves the directivity of the antenna system, thereby increasing the field of view. Along similar direction, a equiangular spaced four monopole configuration can also be considered, as shown in \figurename \ref{fig:antennas}(d). However, the number of correlations is much higher and consequentially demanding more signal processing hardware for each antenna. In the OLFAR study where a $3U$-cubesat ($30\times 10\times 10$ cm) is utilized, the monopoles are deployed in groups of three at the opposite ends of the satellite, as seen in \figurename \ref{fig:antennas}(e). The asymmetric design changes the properties of the monopoles and reduces the purity of the independent components, which is studied by \citet*{smith13} and experimentally evaluated on a smaller scale by \citet*{quillien2013}.

%%-----------------------------------------------------------------------------------------------------------------------%

\begin{figure}[t!] %\sidecaption
    \includegraphics[width=0.6\textwidth]{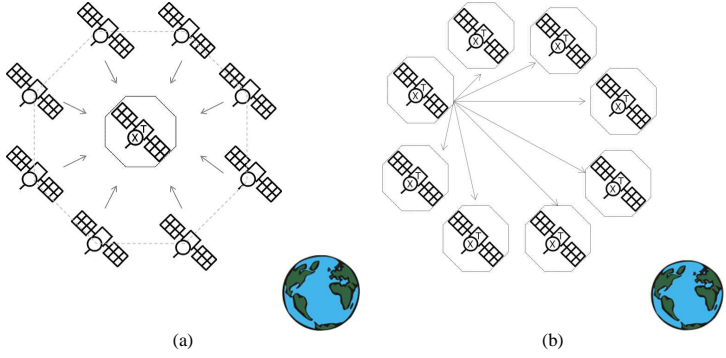}
  \caption{\emph{\textbf{Correlator architectures:}} \footnotesize An illustration of two potential correlator architectures for space-based radio interferometric array, where the tags `X' and `T' on the nodes indicate Correlation and Transmission to Earth capabilities respectively. In the \textbf{(a) Centralized correlator} architecture a centralized mother ship receives data from all observational satellites, correlates and down-links data to Earth. On the contrary, in the \textbf{(b) Distributed correlator} framework, the observed data is evenly distributed between all nodes. Post correlation, all satellite nodes down-link their respective correlated data to Earth. }
  \label{fig:Correlators}
\end{figure}

\section{Digital Signal Processing}  \label{sec:processing} Sky images in radio astronomy are made by calculating the fourier transform of the measured coherence function \citep{taylor1999}. The coherence function is the cross correlation product between two antenna signals located at the two spatial positions, averaged over a period of the integration time $\tau_{int}$. One way to implement such a system is using the traditional XF correlator $\ie$ cross correlation first and fourier transform later and the more recent FX correlator which measures directly the cross-power spectrum between the two antenna signals \citep{bunton2004}. Although the XF architecture is beneficial because bandwidth can be traded for spectral resolution, FX architecture offers computational efficiency and more significantly. The processing factor for XF vs FX is given by \begin{equation}
N^{xf/fx}_{X}= \Big(\frac{N_{sig}N_{bins}}{N_{sig}+ \log_2 N_{bins}}\Big),
\end{equation} where  $N_{sig}= N_{pol}N$ and $N_{bins}= \Delta f_i/\Delta f_{res}$ \citep{rajan2013ac2}. Observe that the multiplicands in the XF mode are additive in the FX mode , besides the $\log_2$ reduction on the number of frequency bins. Thus, although for lower number of nodes the XF is comparable to FX mode, for large scalable architectures the FX mode is computationally cost effective. Since we prefer a scalable space based array the FX architecture is chosen as the preferred architecture . \tablename\ \ref{tb:dataRatesDSP} shows data rates for cluster of $N=9$ nodes, with instantaneous bandwidth of $\Delta f_i=1$MHz and $\tau=1$ second integration time.

\begin{figure}[t!] %\sidecaption
\centering
\includegraphics[width=0.6\textwidth]{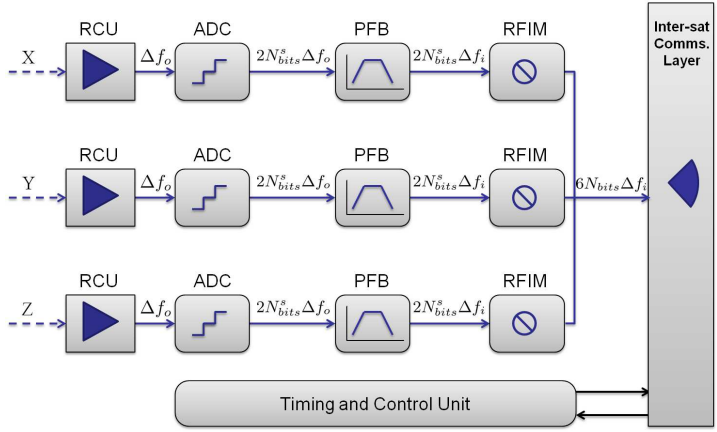}
\caption{\textbf{\emph{Node level signal processing:}} \footnotesize Given the low observational bandwidth of $\Delta f_o \le 50$ MHz, the $N_{pol} =3$ polarized astronomical signals received by each antenna will be signal conditioned and Nyquist sampled at $2\Delta f_o$ MHz with a $14$-bit (or more) Analog to Digital Converter (ADC) as shown in \figurename \ref{fig:figNSP}. A coarse Poly-phase Filter Bank (PFB) is used to selectively choose the desired instantaneous bandwidth of $\Delta f_i= 1$MHz. After successful RFI Mitigation (RFIM), only $N_{bits}=1-2$bits will be used in further processing stages \citep{altunin2001}. The total data generated for $N_{pol}$ signal paths in each satellite is $D_{obs}  =6\Delta f_iN_{bits}$bps, which is transported to the Inter-satellite communication layer.}
\label{fig:figNSP}
\end{figure}

A typical pre-processing unit at each satellite node is shown in \figurename\ \ref{fig:figNSP}, where each satellite generates $D_{obs}  =2\Delta f_iN_{pol}N_{bits}$bps. Observe that with $N_{pol}=3$, for a signal with nominal instantaneous bandwidth of $\Delta f_i= 1$MHz sampled with $N_{bit}= 1$bit resolution, the output data rate is $6$Mbps. Given a far away deployment location, such as Lunar orbit (\ssim $400,000$ km) or Earth leading/trailing (\ssim $2\times 10^6$ to \ssim $4\times 10^6$  km), this down-link data rate levies heavy prerequisites on the limited resources of a small satellite using current techonology. Hence, the satellite cluster must not only employ on-board pre-processing of astronomical signals, but also on-board correlation to minimize down-link data rate back to Earth. To this end, either a centralized or a distributed correlator can be employed as illustrated in \figurename \ref{fig:Correlators}.

\subsection{Centralized architecture} In the centralized FX correlator framework each satellite node transmits $D_{obs}  =2\Delta f_i N_{pol}N_{bits}$bps to the centralized mothership, which in turn receives $D^{c}_{in} =\  D_{obs}(N-1)$bps in total, excluding the data collected from the antenna on the mothership itself. The input data from all satellites is then correlated and the output is then transmitted down to Earth at the rate $D^{c}_{out} =\  2N^2_{sig}N_{bins}/ \tau$bps, where $N_{bins}= \Delta f_i/ \Delta f_{res}$. A significant drawback of the centralized correlation is that it depends heavily on the healthy operation of a single correlation station, which introduces a Single Point Of Failure (SPOF) for large array of satellites in space.

\subsection{Distributed architecture} To alleviate SPOF a \emph{Frequency distributed correlator} is proposed where each node is pre-assigned a specific sub-band $\Delta f_{sb}$ of the observed instantaneous bandwidth $\Delta f_{sb}$ for cross correlations \citep{rajan2013ac2}. Hence, in addition to the node pre-processing (\figurename \ref{fig:figNSP}), a secondary fine PFB is implemented to further split the instantaneous bandwidth $\Delta f_i$ into $N_{sb}$ sub bands, each of bandwidth $\Delta f_{sb} = \Delta f_{i}/N_{sb}$. Each satellite is assigned a specific sub-band for processing and the other $(N_{sb}-1)$ sub-bands are transmitted to corresponding satellites via the intra-satellite communication layer. Furthermore, for even distribution of data and to ensure scalability, we enforce the number of sub-bands equal to the number of satellite nodes $\ie\ N_{sb} =\ N$. Subsequently, in the distributed framework, each node receives a specific sub-band of the observed data \ie $(D_{obs}/N_{sb})$ from $N-1$ other satellites in the network which yields a total input of $D^{d}_{in}=  (D_{obs}/N_{sb}) (N-1) = (D^{c}_{in}/N)$bps, and down-links $D^{d}_{out}= (D^{c}_{out}/N)$bps respectively.

\begin{table*}[!t]
\begin{center}
\begin{tabular}{|l|l|l|l|}
\hline
\textbf{Data rates and processing} 	& \textbf{Notation/Equation}& \textbf{Value} & \textbf{Units}/\textbf{Remark} \\
\hline
No. of satellites (or antennas)  		& $N$   & 9  & (scalable)\\
No. of polarizations             		& $N_{pol}$   & 3 & \\
No. of channels/signals							& $N_{sig}= N_{pol}N$ & 27 & \\
No. of bits                      		& $N_{bits}$   & 1 & bits \\
Observation frequency range         & $\Delta f_o$  & \ssim$30$ &MHz \\
Instantaneous bandwidth             & $\Delta f_{i}$ & 1 &MHz \\
Spectral resolution                 & $\Delta f_{res}$ & 1 & kHz \\
No. of bins													& $N_{bins}=\Delta f_i/ \Delta f_{res}$ & 1000 & \\
Snapshot integration time           & $\tau$ &1 & second \\
%Sub-band bandwidth                  & $\Delta f_{sb}= \Delta f_iN^{-1} $ & 100 & kHz \\
Observed data rate & $D_{obs}= 2\Delta f_i N_{pol}N_{bits}$& 6 & \mbox{Mbps/satellite} \\
\hline
\bf{Centralized} & & &\\
\hline
mothership data reception & $D^c_{in}= D_{obs}(N-1)$& 48 & \mbox{Mbps} \\
Earth down link data rate & $D^c_{out}= 2N_{sig}^2N_{bits}N_{bins}/\tau$& 1.46 & \mbox{Mbps} \\
\hline
\bf{Distributed} & & &\\
\hline
No. of sub-bands										& $N_{sb}=N$													& 9   & 		\\
Sub-band bandwidth                  & $\Delta f_{sb}= \Delta f_i/N_{sb} $ & 111.11 & kHz \\
Inter-satellite reception & $D^d_{in}=  D^c_{in}/N $& 5.34 & \mbox{Mbps/satellite} \\
Earth down link data rate & $D^d_{out}= D^c_{out}/N$ & 162.2 & \mbox{kbps/satellite} \\
\hline
\end{tabular}
\caption{\emph{\textbf{Centralized vs Distributed processing:}} Data rate estimates for a \emph{Centralized correlator} and a \emph{Frequency distributed FX correlator} for the DARIS mission of $9$-satellites.}
\label{tb:dataRatesDSP}
\end{center}
\end{table*}

Thus, the \emph{Frequency distributed correlator} reduces the inter-satellite communication by a factor $N$. Furthermore, at the cost of equipped quipping all observational satellite nodes with communication capability (both Inter-satellite and down-link to Earth), SPOF is averted and scalability is ensured. In the context of the projects discussed earlier, DARIS, FIRST and SURO-LC implement a centralized architecture, whereas OLFAR employed the distributed architecture. Given that the system frequency is typically an order or more than the processing instantaneous bandwidth, computing requirements are negligibly small, which has been duly noted by all these studies.

\subsection{Clocks}  \label{sec:clocks} 
\begin{figure}[t!] %\sidecaption
\centering
\mbox{
\scalebox{0.5}[0.55]{\includegraphics{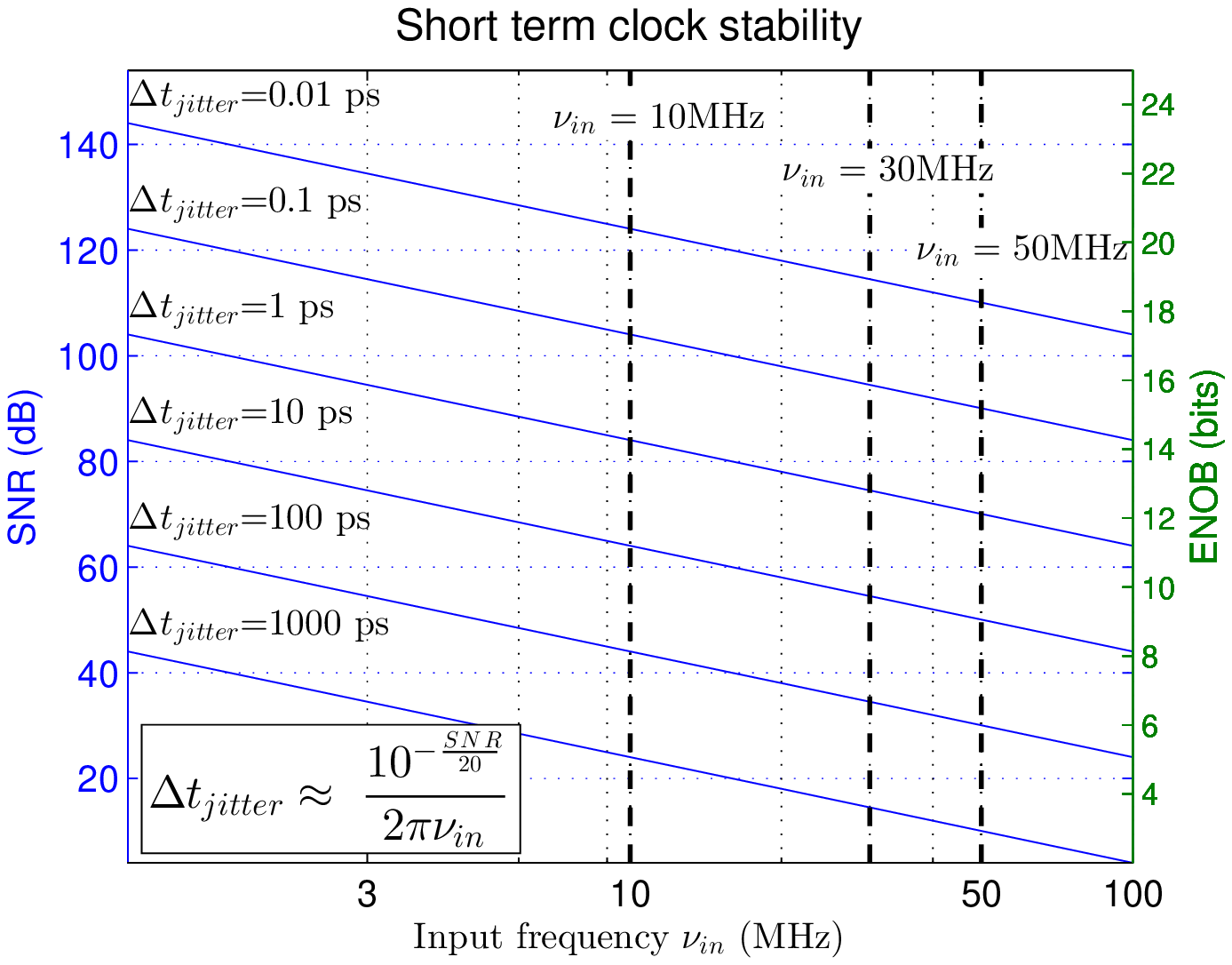}} \hspace{-2.5mm}
\scalebox{0.55}[0.55]{\includegraphics{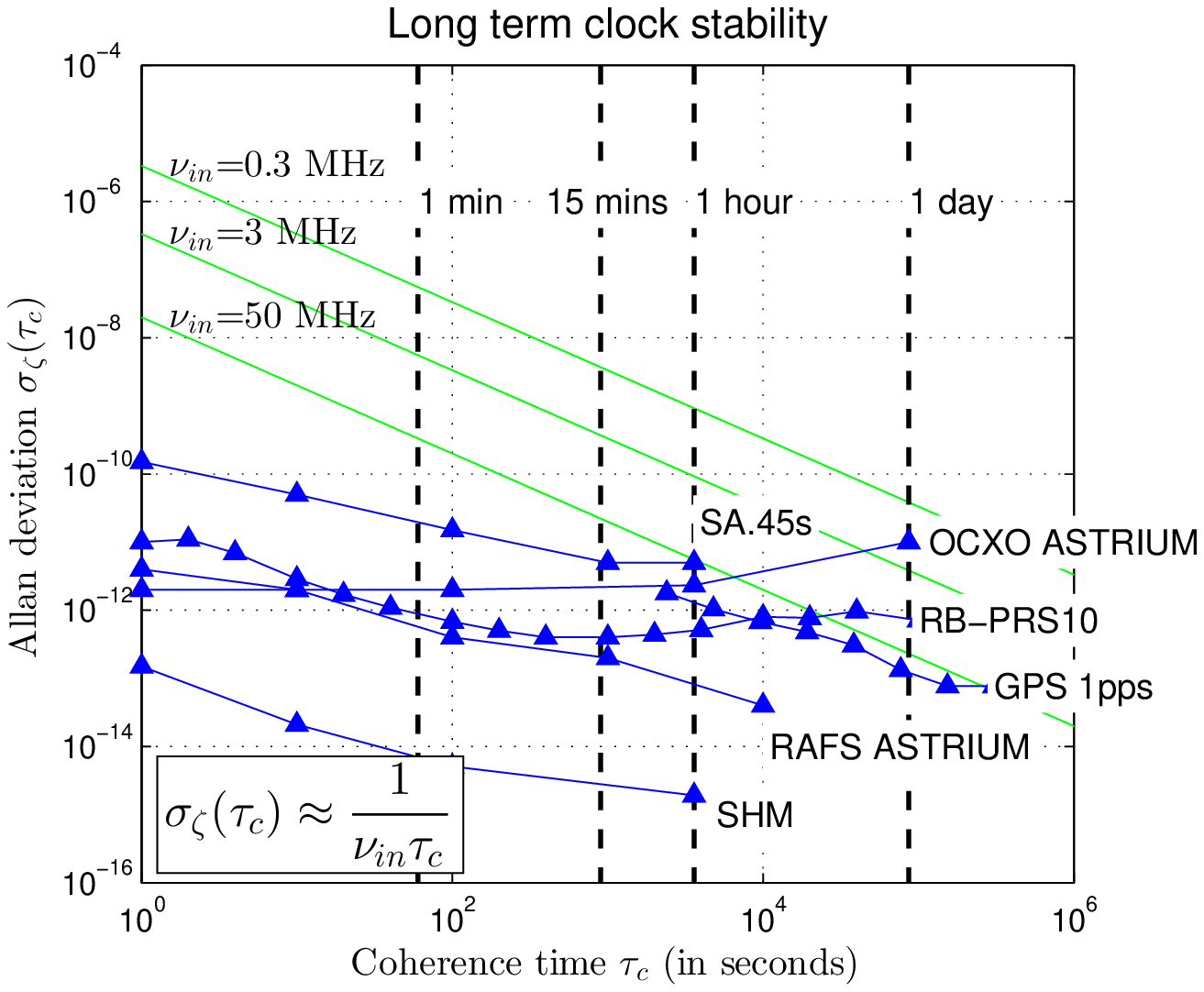 }}
}
\caption{\textbf{\emph{Clock stability}}: \footnotesize (Left) Short term: The plot shows limiting cases of the Signal to Noise Ratio (SNR) and corresponding Effective Number of Bits (ENOB) due to jitter  $t_{\text{jitter}}$ versus input frequency $\nu_{in}$ . Three demarkation lines shows the maximum input frequency of $\nu_{in}=10$MHz, $\nu_{in}=30$MHz and $\nu_{in}=50$MHz. (Right) Long term: Desired Allan deviations of free running clocks are plotted versus the coherence time (in green) for various input frequencies $\nu_{in}$. The map is overlayed with Allan deviations of potential clocks (in blue) for potential space-based low frequency arrays namely PRS-10 Rubidium \protect \citep{standford2006} , RAFS ASTRIUM \protect \citep{droz2007}, GPS 1pps \protect \citep{lombardi2001},  OCXO ASTRIUM \protect \citep{ocxo2012}, SA.45s CSAC \protect and Space Hydrogen Maser (SHM) \protect \citep{goujon2010}.
}
\label{fig:clock}
\end{figure}

The choice of the on-board clock on each satellite has a significant impact on the signal processing system . The short-term clock stability \ie $t \ll 1$ second is dominated by the clock jitter, which limits the Effective Number Of Bits (ENOB) for a given input frequency $\nu_{in}$. For instance, to facilitate $14$-bit sampling at $\nu_{in}$ the chosen clock must have $\delta t_{jitter}<1$ps, as seen in the \figurename\ \ref{fig:clock}(a). Secondly, to evaluate clock stability over long durations $\ie\ t \gg 1$ seconds, we use \emph{Allan variance}, which is a measure of nominal fractional frequency drift \citep{barnes1971}. Following \citep{thompson1994_phaseStability,ulvestad1986}, the stability requirement on the clock and define the \emph{coherence time} $\tau_c$, such that the RMS phase error of the clock remains less than 1 radian \begin{equation} \label{eq:coherenceLimit} 
\nu\sigma_{\zeta}(\tau_c)\tau_{c} \lessapprox 1 ,
\end{equation} where $\nu$ is the observational frequency and $\sigma_{\zeta}(\tau_c)$ is the Allan deviation as a function of $\tau_c$ \citep{rajan2013ac1}. The product $\sigma_{\zeta}(\tau_c)\tau_{c}$ can be visualized as the time drift due to non-linear components of the clock after $\tau_{c}$ seconds. Furthermore, the linear parameters of the clock \ie frequency and phase offsets can be eliminated by exploiting the affine clock model, which is discussed in Section \ref{sec:syncLoc}. \figurename \ref{fig:clock} shows expected Allan deviations of potential clocks versus the coherence time as per (\ref{eq:coherenceLimit}) for various input frequencies $\nu_{in}$. Among the presented choices in \figurename \ref{fig:clock}, the RAFS ASTRIUM ($3.3$kg, $30$W) and OCXO ASTRIUM ($220$g, $2$W) are space qualified Rubidium and Oven controlled oscillators respectively. A particular clock of interest is the SA.45s, which is a Rubidium clock weighing less than $35$ grams, consuming $< 0.125$ watts and offers an coherence time of up to $~15$ minutes.

%%-----------------------------------------------------------------------------------------------------------------------%
\section{Communications}  \label{sec:communication} 
\begin{figure}[t!] %\sidecaption
    \includegraphics[width=0.7\textwidth]{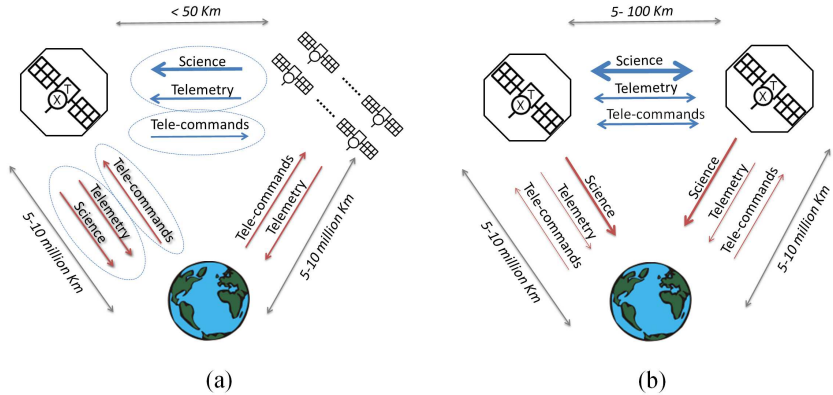}
  \caption{\emph{\textbf{Communication architectures:}} \footnotesize An illustration of a \textbf{(a) Centralized communication architecture} and  \textbf{(b)} single pairwise-link of a \textbf{Distributed communication architecture} for a space-based radio interferometric array. The Inter-satellite link is indicated in \color{blue} \textbf{blue} \color{black} and the Earth-downlink by \color{red} \textbf{red} \color{black}. Telemetry and Tele-commands are exchanged between the satellites and with Earth in both scenarios.}
  \label{fig:Comms}
\end{figure}

The potential communication scenarios for the envisioned space-based array are shown in \figurename \ref{fig:Comms}, which follow directly from the correlator architectures discussed in the previous section. A centralized architecture, as shown in \figurename \ref{fig:Comms} (a), comprises of a mothership collecting raw observed data from a cluster of daughter satellites and, down-links the processed data to an Earth-based ground station. Alternatively, in case of the distributed scenario shown in \figurename \ref{fig:Comms} (b), all satellites are capable of both exchanging data and correlating them, before down-linking back to Earth. In addition to the science data, housekeeping information is also exchanged between the satellites via tele-commands and telemetry, which is expected to be relatively small ($\lessapprox 100$kbps) in comparison to the astronomical data of $6$Mbps.
The housekeeping information is critical for control, timing and synchronization of the satellite, and to maintain coherence within the satellite network.

\subsection{Inter-Satellite Link (ISL)}  \label{sec:isl} Implementing the ISL using high-frequency optical communication \citep{sodnik2006,toyoshima2005} has many advantages as compared to radio communication, such as reduced mass and volume of equipment, higher data rates and no regulatory restrictions as experienced for Radio Frequency (RF) bands. However, this would also require extremely accurate alignment of the satellites, robust synchronization and more power than what could potentially be available for a small satellite. In the RF domain, OFDM is an efficient modulation scheme for the ISL, in particular for a scalable antenna array with limited available bandwidth \citep{nee2000ofdm}. The main advantages of OFDM are it is very well suited to frequency selective channels and it potentially offers a good spectral efficiency. With OFDM, the separation between each channel is equal to the bandwidth of each channel, which is the minimum distance by which the channels can be separated. The signals from each satellite node which form individual channels will be modulated using a form of Phased Shift Keying-PSK, Amplitude Shift Keying-ASK, or a combination Quadrature Amplitude Modulation-QAM. In this article, we consider an ISL transmission frequency of $2.45$GHz, although other frequency bands can also be used.

\begin{figure}[t!] % \sidecaption
  \includegraphics[width=1\textwidth]{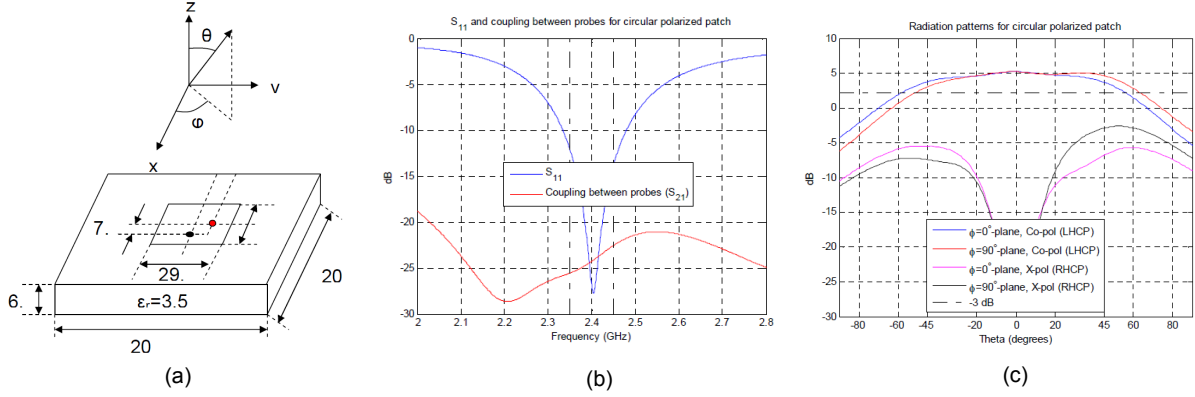}
  \caption{\emph{\textbf{Patch Antenna for Inter-Satellite Link (ISL):}} \footnotesize (a) The simulated patch antenna with dimensions in millimeters. (b) Reflection coefficient and  (c)  Radiation patterns across desired frequency range.}
  \label{fig:patch}
\end{figure}

One of the possible solutions to implement the ISL is to use patch antennas on each face of a satellite node, such that the combined implementation yields a full coverage of the sky. All the satellites will have patch antennas on all six faces for both uplink and downlink. In addition, a diplexer will be used to separate the receiving and transmitting channels. Using a coaxial switch (controlled by the CPU) this signal is selected, amplified and finally detected in the node. The desired antenna must have a bandwidth of $100$MHz around $2.4$GHz with the reflection coefficient less than $-10$ dB between $2.35$GHz and $2.45$GHz, and the half-power beam-width must be at least $90^{\circ}$. \figurename\ \ref{fig:patch}(a) shows the simulated patch antenna, where all dimensions are in millimeters. The patch is fed by a coaxial probe located $7.5$mm from the center of the patch. It was observed that in case of a linearly polarised patch antenna, the given setup yields a $-3$ dB beamwidth less than $90^{\circ}$ for the radiation pattern in the $\phi=90^{\circ}$ plane, which is less than our desired requirement. Thus, to improve the radiation pattern we implemented a circularly polarized patch antenna. Moreover, an added advantage is that the polarization of both transmit and receive antenna is independent of the orientation of the antennas with respect to each other. The circular polarization is realized by adding a second co-axial probe to the patch, shown in red in \figurename\ \ref{fig:patch}(a). As seen in the \figurename\ \ref{fig:patch}(b), the reflection coefficient is better than $-10$ dB in the frequency range $2.35-2.8$ GHz. Additionally, the radiation patterns in \figurename\ \ref{fig:patch}(c) show the half-power beam width is more than $3$dB for both the $\phi=0^{\circ}$ and $\phi=90^{\circ}$ planes.

\subsection{ISL Link margin}  \label{sec:islLinkmargin} \tablename\ \ref{tb:ISLlinkmargin} shows the ISL Link budget for the centralized and distributed scenarios. In case of the centralized scenario, we assume that the mothership is in the center of the array with a maximum mothership-Node distance of $50$km. To estimate the link margin for a centralized scenario, we assume that the mothership can transmit with a power of $1$W \ie $30$dBm. Now, using the patch antenna gain of $3$dBi and nominal losses (at the diplexer, transmitting cable and coaxial switch loss) of $2.6$ dB, the EIRP (Equivalent Isotropically Radiated Power) of each satellite is $30.4$dBm. For both centralized and distributed scenarios, the ISL channel in space is in principle free space loss, where Multi-path, atmospheric losses, absorption losses and even Doppler effects can be ignored. Hence, the Friis free space loss for $2.4$GHz transmission frequency and $50$km is $134.7$dB. Hence, the effective received power for an EIRP of $30.4$dB and a pointing loss of $0.5$dB is $-104.80$dB. The received C/N0 is estimated at $67.66$dB/Hz, for a G/T ratio of $-26.14$ dB/K (see \tablename\ \ref{tb:ISLlinkmargin}). The transmission data rate mothership to Node is $100$kbps ($50$ dB/Hz), which yields an Eb/No of $17.66$dB. For a typical receiver, an Eb/No of $2.5$dB is needed. Now, including a implementation loss of $2$dB, the link margin for uplink with transmit power of $1$W is $13.16$dB. The downlink from the Node satellite to the mothership, includes the $6$Mbps data (see \tablename\ \ref{tb:dataRatesDSP}) and the housekeeping data of $100$kbps, which amounts to $6.10$Mbps. This downlink can be achieved with a link margin of $2.29$ with a transmission power of $5$W.

Extending the link margin estimates of the centralized ISL architecture to a distributed scenario has two fundamental challenges. Firstly, the transmission data rate between every satellite pair is now $5.44$Mbps, which includes $5.34$Mbps of science data (see \tablename\ \ref{tb:dataRatesDSP}) and $100$Kbps of housekeeping data. Secondly, in the absence of a centralized correlator, the maximum distance between the satellites is $100$km, a factor $2$ compared to the centralized scenario. Hence, to achieve the same link margin of $2.29$dB as the Node to mothership downlink, the transmission power of each satellite in the distributed architecture must be $4$ times that of the centralized scenario, \ie $20$W. Although, $15$W suffices to achieve a positive link margin for the distributed architecture. This requirement is a bottle neck for scalable array of small satellites with limited transmission power. One possibility is to use clustering schemes and multi-hop approaches to reduces the communication distances between the satellites \citep{budianu2011iac}, which is a research area currently being explored \citep{naghshvar2012}.

\begin{table}[t]
\resizebox{\textwidth}{!}{ \tiny
\begin{tabular}{|l|l|ll|l|l|} \hline
                  &
                  & \multicolumn{2}{c}{\textbf{Centralized}} \vline
                  & \multicolumn{1}{c}{\textbf{Distributed}} \vline
                  & \multicolumn{1}{c}{\textbf{}} \vline \\
                  & \multicolumn{1}{l}{\textbf{Parameters}} \vline
                  & \textit{MS $\rightarrow$ Node}
                  & \textit{Node $\rightarrow$ MS}
                  & \textit{Node $\leftrightarrow$ Node} & \multicolumn{1}{l}{\textbf{Units}} \vline \\ \hline
System Definition & Frequency band                          & 2.45                           & 2.45                           & 2.45                                      & GHz                                \\
                  & Maximum distance                        & 50.00                          & 50.00                          & 100.00                                    & km                                 \\
                  & TX power                                & 1.00                           & 5.00                           & 15.00                                     & W                                  \\
                  & Required datarate                       & 0.10                           & 6.10                           & 5.44                                      & Mbps                               \\
                  & Antenna gain                            & 3.00                           & 3.00                           & 3.00                                      & dBi                                \\
                  & Wavelength                              & 0.12                           & 0.12                           & 0.12                                      & m                                  \\ \hline
ISL Channel       & Free space Path losses                  & 134.20                         & 134.20                         & 140.23                                    & dB                                 \\
                  & Atmospheric losses                       & 0.00                           & 0.00                           & 0.00                                      & dB                                 \\
                  & Polarization losses                     & 0.50                           & 0.50                           & 0.50                                      & dB                                 \\
                  & Absorption losses                       & 0.00                           & 0.00                           & 0.00                                      & dB                                 \\
                  & Path loss                               & 134.70                         & 134.70                         & 140.73                                    & dB                                 \\ \hline
Link Budget       & TX power                                & 30.00                          & 36.99                          & 41.76                                     & dBm                                \\
                  & Loss coaxial switch                     & 0.30                           & 0.30                           & 0.30                                      & dB                                 \\
                  & Diplexer losses                         & 1.00                           & 1.00                           & 1.00                                      & dB                                 \\
                  & Cable losses                            & 1.30                           & 1.30                           & 1.30                                      & dB                                 \\
                  & Transmit antenna gain                   & 3.00                           & 3.00                           & 3.00                                      & dBi                                \\
                  & EIRP of the spacecraft                  & 30.40                          & 37.39                          & 42.16                                     & dBm                                \\
                  & Pointing loss                           & 0.5                            & 0.5                            & 0.5                                       & dB                                 \\
                  & Received power                          & -104.80                        & -97.81                         & -99.06                                    & dBm                                \\
                  & Receiver G/T                            & -26.14                         & -26.14                         & -26.14                                    & dB/K                               \\
                  & Bolzmann constant                       & -198.60                        & -198.60                        & -198.60                                   & dBm/Hz/K                           \\
                  & Received C/N0                           & 67.66                          & 74.65                          & 73.40                                     & dB/Hz                              \\
                  & Data rate                               & 50.00                          & 67.85                          & 67.36                                     & dB/Hz                              \\
                  & Received Eb/No                          & 17.66                          & 6.79                           & 6.04                                      & dB                                 \\
                  & Required Eb/No                          & 2.5                            & 2.5                            & 2.5                                       & dB                                 \\
                  & Implementation loss                     & 2                              & 2                              & 2                                         & dB                                 \\ \hline
                  & Link Margin                             & 13.16                          & 2.29                           & 1.54                                      & dB \\ \hline
\end{tabular} } 

\vspace{5mm}

\resizebox{\textwidth}{!}{
\begin{tabular}{|l|llllllll|}
\hline
\emph{\textbf{G/T estimates}} & \textbf{Gain (dB)}        & \textbf{Temp (K)} & \textbf{Temp (dBK)} & \textbf{Loss (dB)} & \textbf{Tref (dBK)}                                  & \textbf{Gref (dB)}  & \textbf{Tref  (K)} &                 \\ \hline
Antenna                   & 3.00              & 100.00              & 20.00              & -2.40                                                & 17.60               & 3.00               & 57.54           &                 \\
Coaxial Cable             & -1.30             & 75.02               & 18.75              & -1.10                                                & 17.65               & -1.30              & 58.23           &                 \\
Diplexer                  & -1.00             & 59.64               & 17.76              & -0.10                                                & 17.66               & -1.00              & 58.29           &                 \\
Coaxial Cable             & -0.10             & 6.60                & 8.20               & 0.00                                                 & 8.20                & -0.10              & 6.60            &                 \\
LNA                       & 30.00             & 288.63              & 24.60              & 0.00                                                 & 24.60               &                    & 288.63          &                 \\
Backend                   & 30.00             & 2610.00             & 34.17              & -30.00                                               & 4.17                &                    & 2.61            &                 \\
LNA Noise Figure (dB)     & 3.00              &                     &                    & \multicolumn{1}{c}{\multirow{3}{*}{\textbf{Totals}}} & \textbf{G (dB)}     & \textbf{0.60}      & \textbf{}       & \textbf{}       \\
Backend Noise Figure (dB) & 10.00             &                     &                    & \multicolumn{1}{c}{}                                 & \textbf{T (K)}      & \textbf{}          & \textbf{471.90} & \textbf{}       \\
Ambient Temperature (K)   & 290.00            &                     &                    & \multicolumn{1}{c}{}                                 & \textbf{G/T (dB/K)} & \textbf{}          & \textbf{}       & \textbf{-26.13} \\ \hline
\end{tabular}
}
\caption{\emph{\textbf{Inter-sallite communication:}} (Above) The Inter-Satellite Link (ISL) budget for centralized and distributed scenarios for a satellite array. (Below) Antenna gain to noise tempearature (G/T) estimates.}
\label{tb:ISLlinkmargin}
\end{table}

\subsection{Space to Earth Downlink} \label{sec:earthDownlink} 
\begin{figure}[t!] %\sidecaption
  \includegraphics[width=0.55\textwidth]{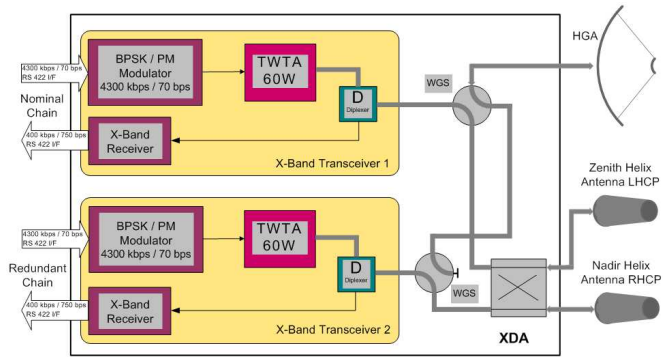}
  \caption{\emph{\textbf{DARIS mothership-Earth Link X-band Downlink Unit (XDU):}} \footnotesize The XDU unit consists of a complete redundant transmit and receive subsystem, the X-band antennas and the interconnecting waveguide links. The X-Band downlink transceivers are directly interfaced   to the Onboard Computer (OBC) via RS422. Two different modes are available for the communication link (a)Nominal link via High Gain Antenna (HGA) for science telemetry and commanding (b) Emergency link via low gain antennas (LGA) with full coverage and reduced data rates for telemetry and command.}
  \label{fig:xdu}
\end{figure}

The total downlink data rate $D_{out}$ after correlation quadratically increases with the number of nodes in the cluster and reduces linearly with the integration time (see \tablename\ \ref{tb:dataRatesDSP}). For a cluster of $9$ satellites, with $1$ second integration interval this rate is $\ge 1.46$Mbps. In the DARIS study, the space to earth downlink was achieved using an X-band Downlink Unit (XDU). \figurename\ \ref{fig:xdu} shows the X-Band transmit chain which consists of a modulator and a $60$W Traveling Wave Tube Amplifiers (TWTA) with $60$\% efficiency. In the nominal operation case the data is directly BPSK modulated on the X-band carrier at $4.3$Mbps, amplified, filtered at the diplexer and transmitted via the high gain antenna ($40$dBi). During launch and early orbit phase and in contingency cases the tranmission rate is reduced to $100$bps, BPSK-modulated onto a subcarrier with low frequency (\eg $8$kHz) which itself PM-modulates the carrier. In order to enable full coverage two adversely polarized low gain antennas are transmitting into the two hemispheres. The receiver downconverts and demodulates the X-band data and feeds them to the OBC via RS422 link. In the nominal mode the command rate is $400$kbps, BPSK-modulated on a $16$kHz subcarrier and PM modulated onto the carrier. The estimated power consumption of the transceiver is $110$W operational and $30$W in standby. The HGA, covering both up- and downlink band, could be realized as a parabolic antenna similar to the one of Mars Express. With a $40$dBi design, the $3$dB beamwidth would be around $1.7^\circ$ posing no big challenge to the pointing mechanism. The LGA is a conventional helix antenna design covering also both transmit and receive band. The number of ground stations on Earth are limited and are therefore almost always in high demand. For instance, the core ESA network has two deep space network $35$m antennas along with several antennas in the $13-15$m class \citep{vassallo2007esa}. A baseline for ground station access for is ESAs $35$m ground station for $8$hours/day suffices the need for the data generated cluster of $10$ satellites with a centralized mothership. For arrays larger than $50$ satellites with minimal power, it is difficult to establish Earth-based downlink with current off the shelf technology. However, these challenges and possible distributed downlink scenarios are currently being investigated \citep{budianu2014acta}.

%One possibility is that the satellites can feature phased-array antenna panels when can be directed towards Earth for downlink \citep{rajan2011ac}. Furthermore, when analyzing the down-link for the various orbital positional scenarios, the Earth-leading and Earth trailing options is singled out, as its a tough challenge to maintain a data rate above the required 900 Kbit/s. In which case, a 24 GHz link will be required, as it provides much more link capacity, at the expense of a much higher complexity. For the Earth-moon link a 5.8 GHz link is sufficient, greatly simplifying the systems complexity and at 24 GHz, a smaller antenna size is acceptable. The Earthmoon L2 case is a special one, as separate relay elements are required to transfer data from the back-side of the moon towards Earth.

%%-----------------------------------------------------------------------------------------------------------------------%
\section{Synchronization and Attitude determination} \label{sec:syncLoc} To maintain coherence between satellites, all the satellites must be synchronized, and their positions known up to sub-meter accuracies. It is worth noting that these requirements on space-time accuracies are much lower compared to other space-based array missions, \eg LISA \citep{bik2007lisa}. Almost all Earth-based antenna arrays synchronize using GPS-aided atomic clocks, where fixed antenna positions are known up to millimeter accuracies, \cf \eg LOFAR \citep{vanHaarlem2013}. However, the envisioned space-based array will be deployed far-away from Earth-based GPS satellites and unlike Earth-based antennas, these satellites will be mobile within a formation or swarm due to orbital dynamics. In addition, given the large number of satellites and limited ground-segment capability, tracking each satellite independently is infeasible. Moreover, in certain deployment locations such as the lunar orbit, the satellite array will be partially or even completely disconnected from Earth-based ground stations during eclipse. 

Hence, the satellite array must be a \emph{cooperative wireless network}, collaboratively estimating time-varying positions and correct for respective clocks simultaneously. More specifically, the satellites will estimate relative positions which suffice for Radio astronomy imaging, inter-satellite communication and collision avoidance. The relative positions can be estimated from distances via MDS-like algorithms \citep*{borg97} \citep*{rajanJ2}. In this section we estimate the clock parameters and the time-varying distances of the network. We are primarily interested in solutions for \emph{cold start} scenarios, when no prior information is known. For longer time scales, when the orbital dynamics of the deployment location is well known, the estimated space-time parameters can be tracked and improved using recursive filters \eg Kalman Filter \citep{kay1993}. The pointing direction for the satellites can be provided by commercially available sun (or star) trackers, which form part of the Attitude and Orbit Control System (AOCS) in the satellites.

\subsection{Joint ranging and synchronization}  
\begin{figure}[t!] %\sidecaption 
\centering
\scalebox{0.28}[0.25]{\includegraphics{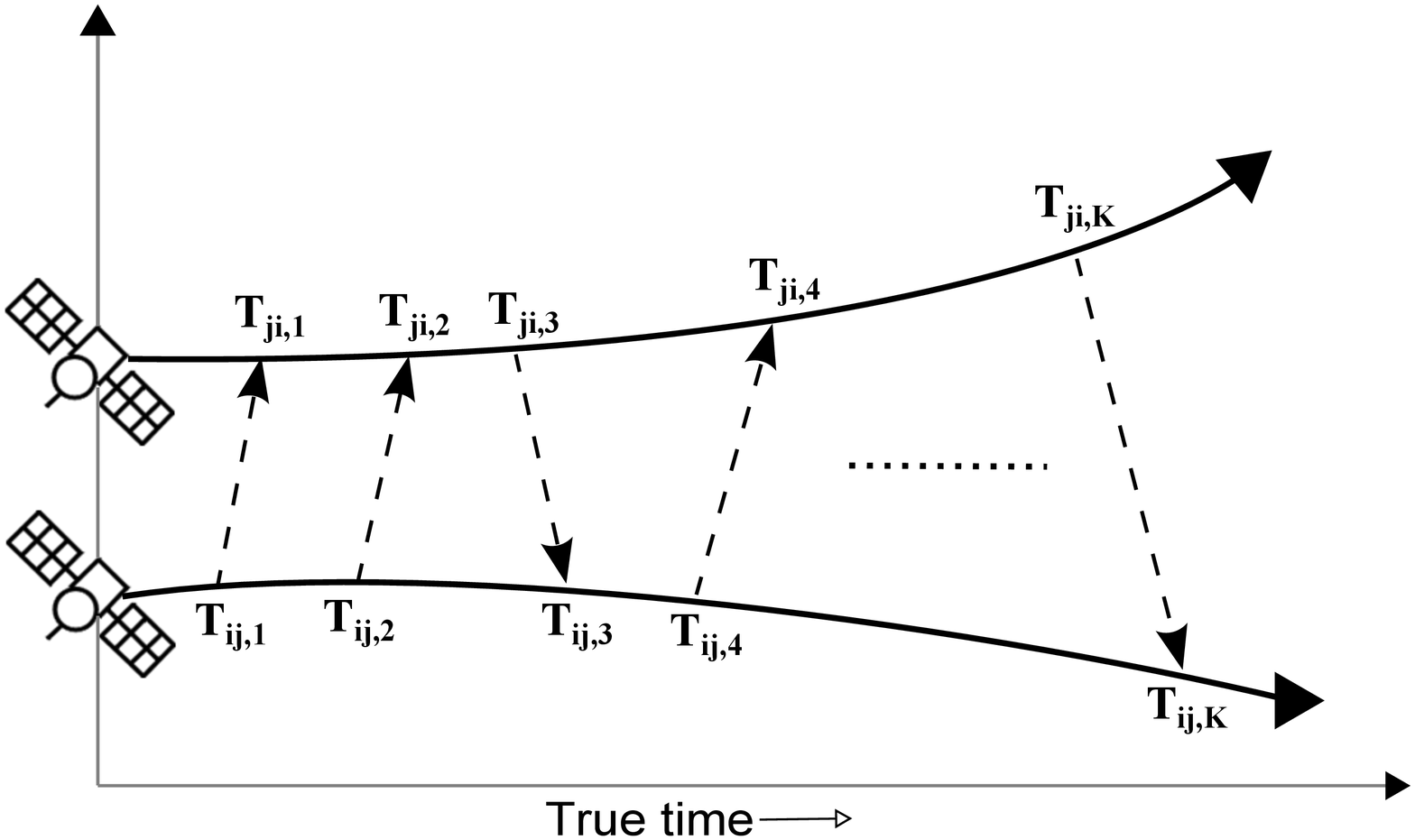}
}
\caption{\textbf{\emph{Dynamic ranging:}} \footnotesize  A generalized Two Way Ranging (TWR) scenario between a pair of \emph{asynchronous mobile satellite nodes} where the asynchronous nodes transmit and receive asymmetrically, during which $K$ time stamps are recorded at respective nodes. Unlike classical TWR \citep{ieee2007} where the transmission and reception is alternating, the proposed setup imposes no pre-requisites on the sequence or number of two way communications. Consequently, this framework and proposed solutions for joint ranging and synchronization \citep{rajanJ1} can be readily extended to a plethora of TWR ranging protocols, including broadcasting and passive listening}
\label{fig:figPairWise}
\end{figure}

All clocks are inherently non-linear \wrt the ideal time $t$. However, a given clock can be approximated to a linear model provided the Allan-deviation of the clock is relatively low for a small coherence time (see Section \ref{sec:clocks}). More generally, let $t_i, t_j$ denote the local times at $i,j$ respectively, then the ideal time $t$ is \begin{eqnarray} \label{eq:clockmodel}
t
\triangleq\ \cC(t_i,\ \dot{\psi}_i,\ \psi_i)
\triangleq\ \cC(t_j,\ \dot{\psi}_j,\ \psi_j),
\end{eqnarray} where  $(\dot{\psi}_i, \dot{\psi}_j)$ and $(\psi_i, \psi_j)$ are the phase and frequency offsets of a satellite pair $(i,j)$ and, $\cC(\cdot)$ represents a linear function of the local clock parameters. However, in case a satellite network, the pairwise distances are time-varying, which can be expanded as \begin{equation}
d_{ij}(t)=    r_{ij} + \dot{r}_{ij}t + \ddot{r}_{ij}t^2 + \hdots,
\end{equation} where $d_{ij}(t)$ is the time-varying distance between the satellite pair $(i,j)$ and $(r_{ij},\dot{r}_{ij},\ddot{r}_{ij}, \hdots)$ are range parameters of the Taylor expansion at \emph{ideal} time $t=0$. More specifically, $r_{ij}$ is the initial pairwise distance, $\dot{r}_{ij}$ is the range rate and $\ddot{r}_{ij}$ denotes the rate of range rate between the satellites. Given these pairwise distances, the relative positions of the satellites can be estimated using MDS-like algorithms\citep{borg97}. Our aim is to jointly estimate the clock parameters and the time-varying pairwise distances between the satellite nodes. The joint synchronization and ranging problem can be formulated as shown in \figurename\ \ref{fig:figPairWise}, which shows a pair of \emph{asynchronous mobile satellite nodes}. The mobile satellites transmit messages asymmetrically between each other, during which $K$ time-markers are recorded at each end. Let $T_{ij,k}$ and $T_{ij,k}$ be the $k$th time-markers recorded at the $i$th and $j$th satellite nodes respectively, and $E_{ij,k} \in \{+1,-1\}$ indicates the transmit and receive direction of the message. For any $k$th time instant, the Generalized Two-Way Ranging (GTWR) equation \citep{rajanJ1} is \begin{equation}
\label{eq:timeRangeBasis}
\cC_i(T_{ij,k}, \dot{\psi}_i, \psi_i) -
\cC_j(T_{ji,k}, \dot{\psi}_j, \psi_j) +
E_{ij,k}d_{ij}\Big(\cC_i(T_{ij,k}, \dot{\psi}_i, \psi_i)\Big)   = 0
\qquad\ \forall i,j \le N\ , \forall k \le K,
\end{equation} where without loss of generality the \emph{ideal} time $t$ of the time-varying distance $d_{ij}(t)$ is replaced with the clock model at satellite $i$ (\ref{eq:clockmodel}). The unknown frequency offsets, phase offsets and the pairwise distances over $K$ time instances can be estimated using the iterative Mobile Pairwise Least Squares (iMPLS) and iterative Mobile Global Least Squares(iMGLS). The iMPLS algorithm is applicable when the daughter satellite nodes communicate only with a centralized Mother-ship (\ie Star network topology, see \figurename\ \ref{fig:Correlators} (a)), whereby only the clocks of the satellites can be corrected. For a full mesh network, when all satellites communicate with each other (\ie Full mesh topology, see \figurename\ \ref{fig:Correlators} (b)) both the clock parameters and the distances can be estimated, which is achieved by the iMGLS algorithm.

\subsection{Simulations} 
\begin{figure*}[t!]%
\centering
\psfrag{sT}[cc]{\small Frequency offset, $\dot{\bpsi}$}
\psfrag{sX}[cc]{\small No. of communications, $K$}
\psfrag{sY}[cc]{\small RMSE [Hz]}
\psfrag{sLa}[lb]{\tiny LCLS}
\psfrag{sLb}[lb]{\tiny iMPLS: Star network}
\psfrag{sLc}[lb]{\tiny iMGLS: Mesh network}
\psfrag{sLd}[lb]{\tiny \Cramer\ Rao Lower Bound}
\psfrag{sLe}[lb]{\tiny Noise on Time-markers}
\psfrag{oT}[cc]{\small Phase offset, $\bpsi$}
\psfrag{oX}[cc]{\small No. of communications, $K$}
\psfrag{oY}[cc]{\small RMSE [seconds]}
\psfrag{oLa}[lb]{\tiny LCLS}
\psfrag{oLb}[lb]{\tiny iMPLS: Star network}
\psfrag{oLc}[lb]{\tiny iMGLS: Mesh network}
\psfrag{oLd}[lb]{\tiny \Cramer\ Rao Lower Bound}
\psfrag{oLe}[lb]{\tiny Noise on Time-markers}
\psfrag{dT}[cc]{\small Distance, $\bd$}
\psfrag{dX}[cc]{\small No. of communications, $K$}
\psfrag{dY}[cc]{\small RMSE [meters]}
\psfrag{dLa}[lb]{\tiny No clock correction}
\psfrag{dLb}[lb]{\tiny After clock correction}
\psfrag{dLc}[lb]{\tiny  \Cramer\ Rao Lower Bound}
\psfrag{rT}[cc]{\small Range parameters, $\br$}
\psfrag{rX}[cc]{\small No. of communications, $K$}
\psfrag{rY}[cc]{\small RMSE }
\psfrag{rLa}[lb]{\tiny Range $\br_{ij}$ [m]}
\psfrag{rLb}[lb]{\tiny Range rate $\dot{\br}_{ij}$ [m/s]}
\psfrag{rLc}[lb]{\tiny Rate of Range rate $\ddot{\br}_{ij}$ [m/s$^2$]}
\psfrag{rLd}[lb]{\tiny  \Cramer\ Rao Lower Bound}
\parbox{2in}{\includegraphics[scale=0.47]{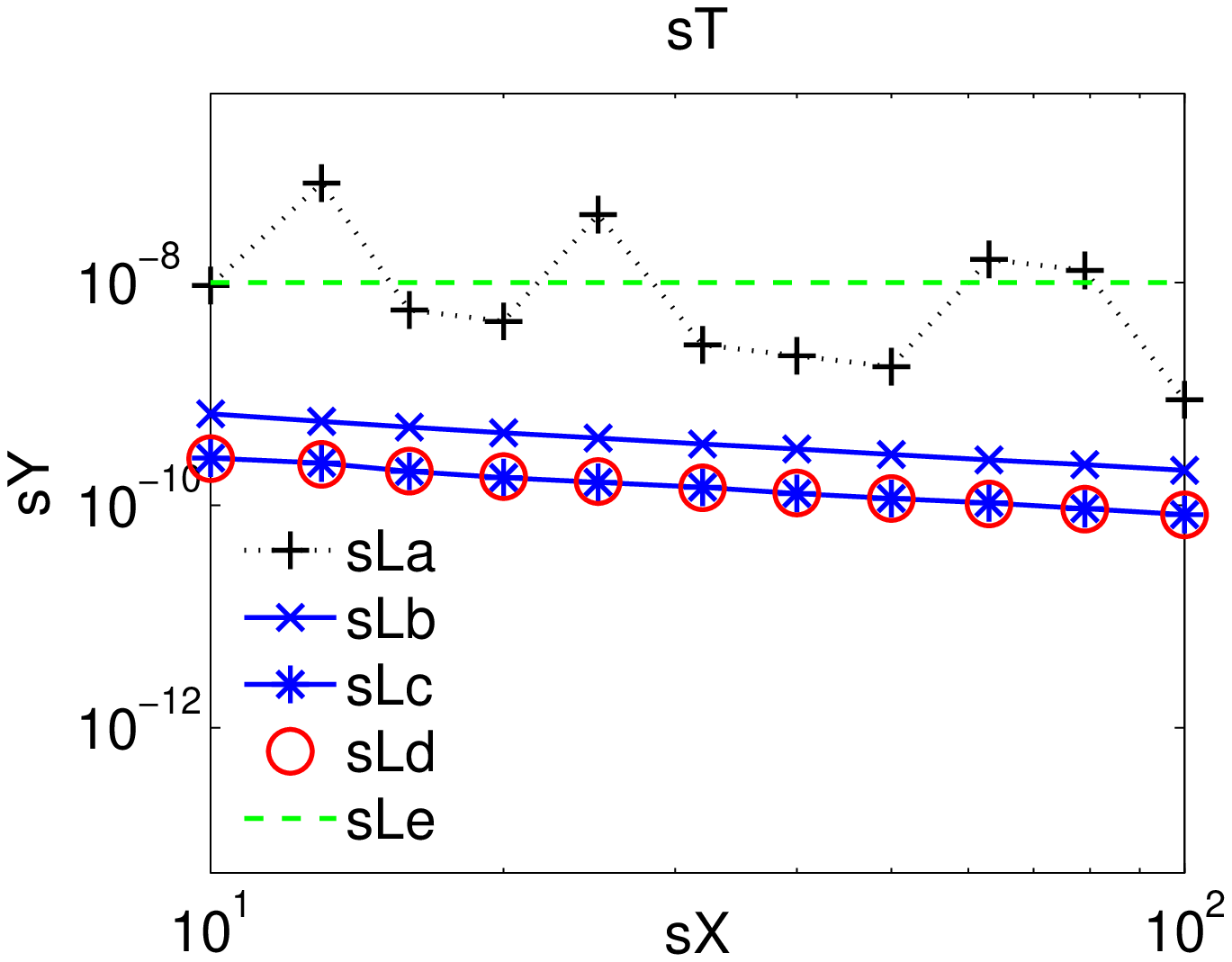}} \hspace{20mm}
\parbox{2in}{\includegraphics[scale=0.47]{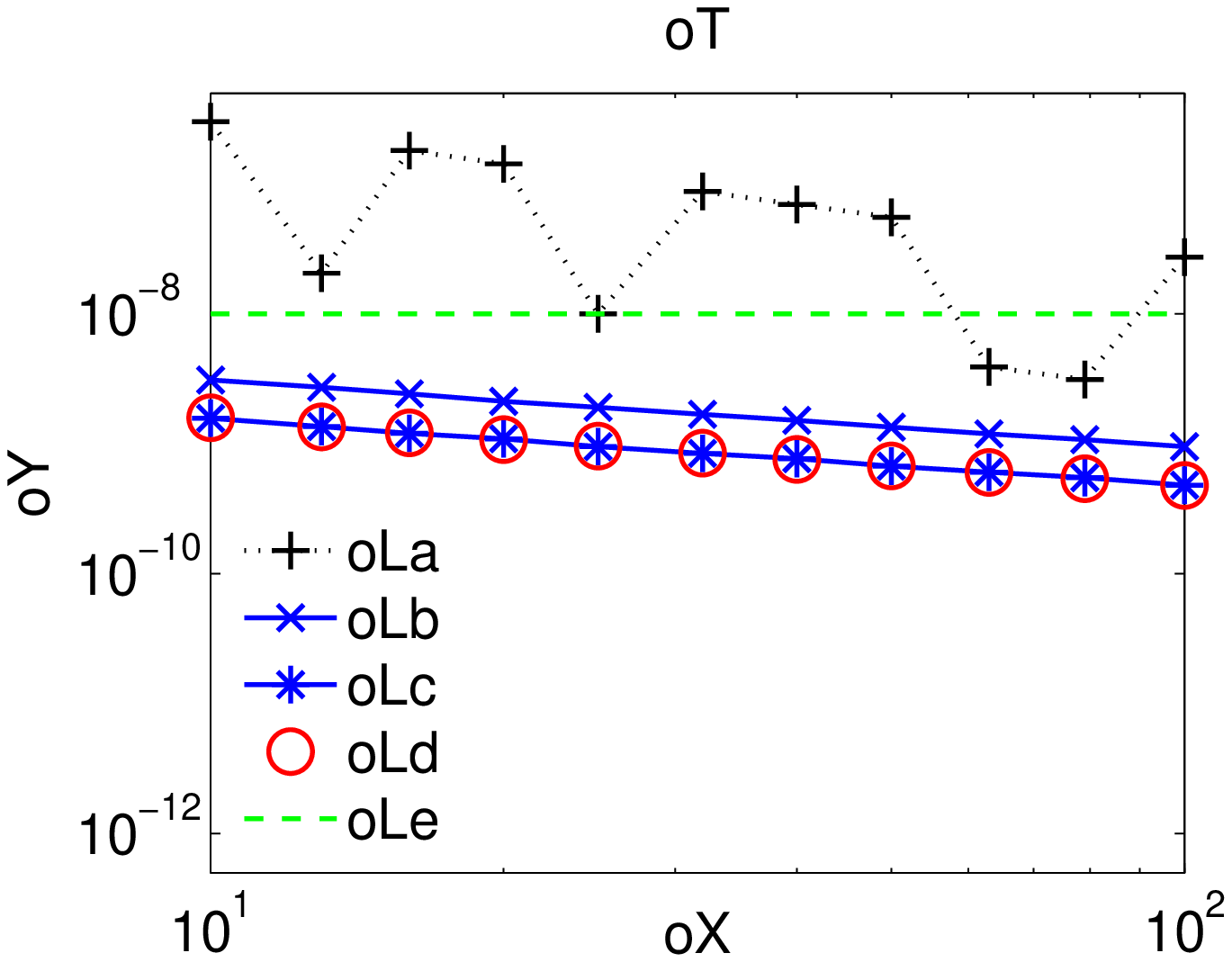}} \\ \vspace{2mm}
\parbox{2in}{\includegraphics[scale=0.47]{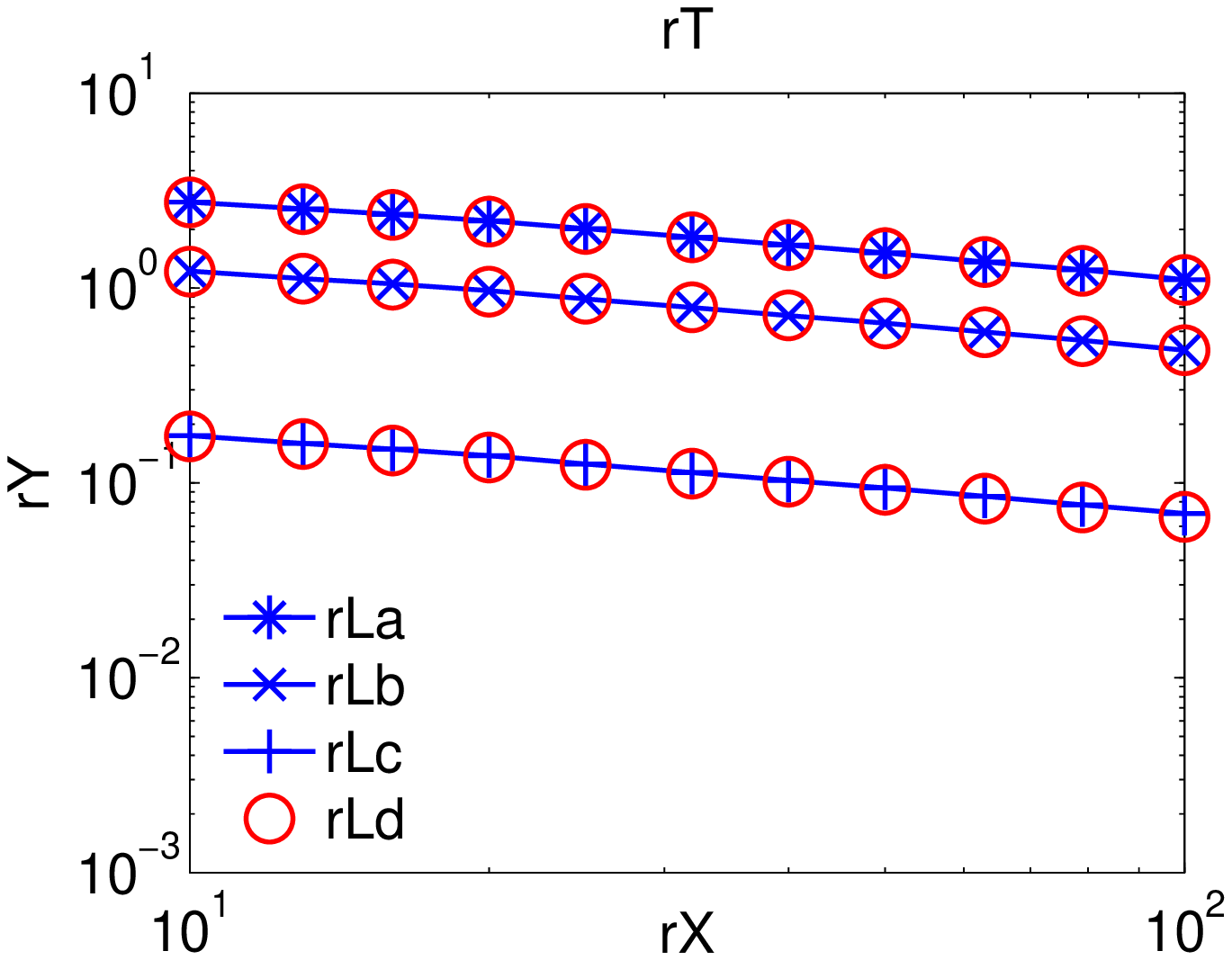}} \hspace{20mm}
\parbox{2in}{\includegraphics[scale=0.47]{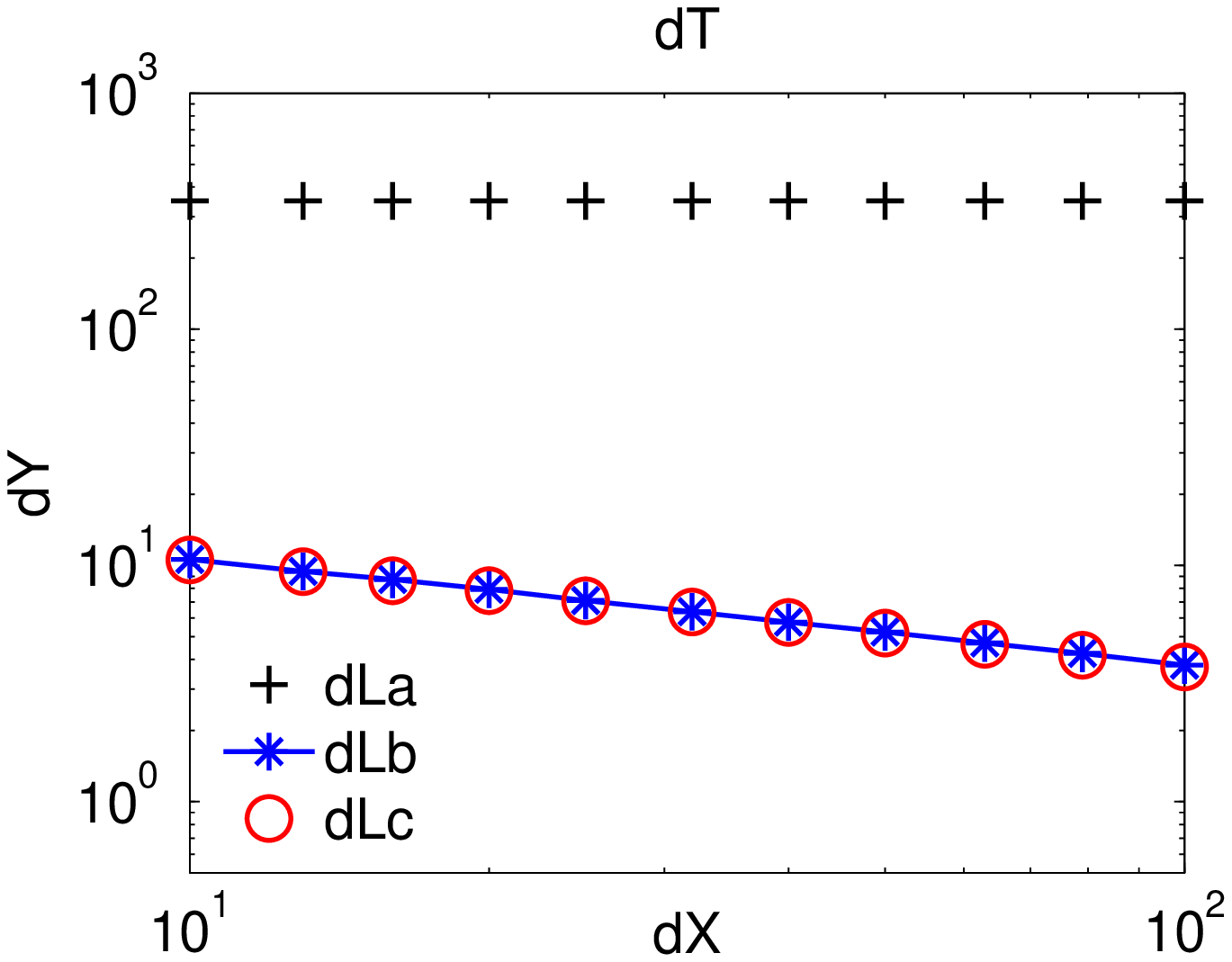}}
%\nocaption
\caption{\emph{\textbf{Joint ranging and synchronization:}} A simulation showing the Root Mean Square Errors (RMSEs) of the estimated frequency offset ($\hat{\dot{\bphi}}$), phase offset ($\hat{\bphi}$), range parameters ($\hat{{\br}}, \hat{\dot{\br}}, \hat{\ddot{\br}}$) and distance $\hat{\bd}$ using the MGLS algorithm for joint ranging and synchronization.}
\label{fig:jointBasisSimulation}%
\end{figure*}

To illustrate the algorithms, we consider a cluster of $N=9$ mobile nodes in a $3-$dimensional Euclidean space. The initial frequency offsets $\dot{\bpsi}=[\dot{\psi}_1, \dot{\psi}_2, \hdots, \dot{\psi}_{9}]$ and phase offsets $\bpsi=[\psi_1, \psi_2, \hdots, \psi_{9}]$ of the nodes are arbitrarily chosen in the range $[-10^{-4}, 10^{-4}]$ and  $[-1, 1]$ seconds respectively, where without loss of generality satellite node $1$ is chosen as the reference node with $[\dot{\psi}_1, \psi_1]= [0,0]$. Secondly, the initial positions $\bX$ and initial velocities $\bY$, whose values are arbitrary chosen as  \begin{eqnarray}
\bX &=&
\begin{bmatrix}[c]
    6.1&   -7.6&   -0.2&    8.2&    9.0&   -8.9&    9.9&    6.0&   -9.3\\
   -1.3&    4.4 &   2.9&   -9.7&   -1.8&   -7.7&   -7.6&    7.8&    7.8 \\
    3.5&    6.8 &  9.1&    5.3&     5.4&    6.6&    6.0&    4.6&    7.5
\end{bmatrix} \text{km}, \nonumber\\
\bY &=&
\begin{bmatrix}[c]
   -70&   -50&   -30&   -70&   -70&    40&   -40&    60&    70\\
    90&    40&    40&   -90&  -100&    10&   -80&   -40&    70\\
    60&    50&    20&   50&     80&    60&    100&    90&   10
\end{bmatrix} \text{ms}^{-1},
\end{eqnarray} For these assumed positions, the initial pairwise distance $\br$ is in the range $[0, 10]$km. Secondly, for these values the range rates $\dot{\br}$ are distributed in the period $[-100, 100]$m/s and the rate of range rate $\ddot{\br}$ are distributed over $[-10, 10]\text{m}^2$/s.  These values are typically for worst case scenarios, assumed to evaluate the performance of the proposed algorithms. The mobile satellites communicate $K$ messages with each other within a time interval of $[-10, 10]$seconds, and the time-markers $T_{ij,k} \forall\ i,j \le N, \forall\ K$ are generated accordingly. We assume alternating communication between the nodes and a Gaussian noise of $\sigma= 10^{-8}\ \text{seconds} \ssim 3.3\ \text{meters}$ plaguing the time-markers. \figurename\ \ref{fig:jointBasisSimulation} shows the performance of the proposed algorithms for phase offset, frequency offset and pairwise distance estimation. The Root Mean Square Errors are plotted (in blue) against varying $K=10$ to $K=100$, where the clock parameters are averaged over $N$ nodes and, the distances and range parameters are averaged over $N \choose 2$ unique links. In addition, the \Cramer\ Rao Lower Bounds for the corresponding estimates are also plotted (in red), which is achieved asymptotically by the proposed estimators. To show the performance of the prevalent solutions, we also plot the Low Complexity Least Squares (LCLS) which synchronizes the clocks for immobile network of satellite nodes. The iMGLS algorithm outperforms the iMPLS estimator since it exploits the full mesh communication network between the nodes. More significantly, the iMGLS estimator achieves clock accuracies upto nanoseconds and distance errors up to meter accuracies at \emph{cold start}.

An added advantage of using dynamic ranging is that the timestamps can potentially piggyback on the housekeeping data exchanged between the satellite nodes, which mitigates the need for a dedicated ranging system. However, if a ranging system is employed, then the achievable lower bound on the standard deviation for Time Of Arrival in multipath-free channels is given by \begin{equation} 
\sigma\ge \big(8\pi^2F^2_cBT\ \text{SNR}\big)^{-1/2}, %\frac{1}{\sqrt{(8\pi^2F^2_cBT\ \text{SNR})}}, 
\end{equation} where $F_c$ denotes the carrier frequency, $B \ll F_c$ is the bandwidth of the signal, $T$ is the signal duration in seconds  \citep{patwari05} . The assumed noise variance on the time-markers in the simulation is $\sigma=10^{-8}$ (shown in green in \figurename\ \ref{fig:jointBasisSimulation}), which can be adequately achieved by a wireless node communicating at $F_c=2.4$GHz with a nominal bandwidth of $1$kHz transmitting and SNR=$10$dB for a signal duration of $T \ssim 1$ms.

%%-----------------------------------------------------------------------------------------------------------------------%
\section{Deployment Locations} \label{sec:deploymentLocations} The deployment location of the space-based array must be chosen to ensure the following conditions. 
\begin{itemize}
\item Minimize RFI during scientific observation cycles.
\item Offer maximum possible down-link data rate.
\item Provide sufficient positional stability during integration time $\tau$.
\item Must remain within a sphere of \ssim$100$km.
\end{itemize} In addition, each satellite must offer low noise conditions with minimal EMC and stable temperature (and gain) conditions to easen calibration easier. To alleviate the high complexity of active control to keep all the satellites within a cluster, \emph{passive formation flying} could be employed. In passive formation flying paradigm, the satellites are allowed to drift, during which the relative positions and orientations of the satellites are constantly monitored. This approach eliminates the need for excess propulsion and heavy orbital maintenance equipment on all satellites. Additionally, the naturally varying position vectors of the satellites produce unique uvw sampling points, which consequentially improve the PSF.

In order to avoid interference either the cluster must be located far from Earth-based RFI and ionospheric distortions, such as Earth leading/trailing orbits and Sun-Earth Lagrangian points. However, by increasing the distance to the Earth, the communication with the Earth becomes more difficult. Alternatively, RFI shielding can be achieved by positioning the array on the far side of the moon. The Radio Astronomy Explorer RAE-2 showed that interference at very low frequencies is reduced by $2$ orders of magnitude behind the moon, making it an ideal location for radio astronomy observations \citep{alexander1975}. However, during the eclipse behind the moon the cluster has no communication with Earth. The following section discusses the quest to find an optimum balance between down-link data rates and maximizing observation time, emphasizing the challenges in various deployment locations.

\begin{figure*}[!t]
\centering
\parbox{2in}{\includegraphics[scale=0.10]{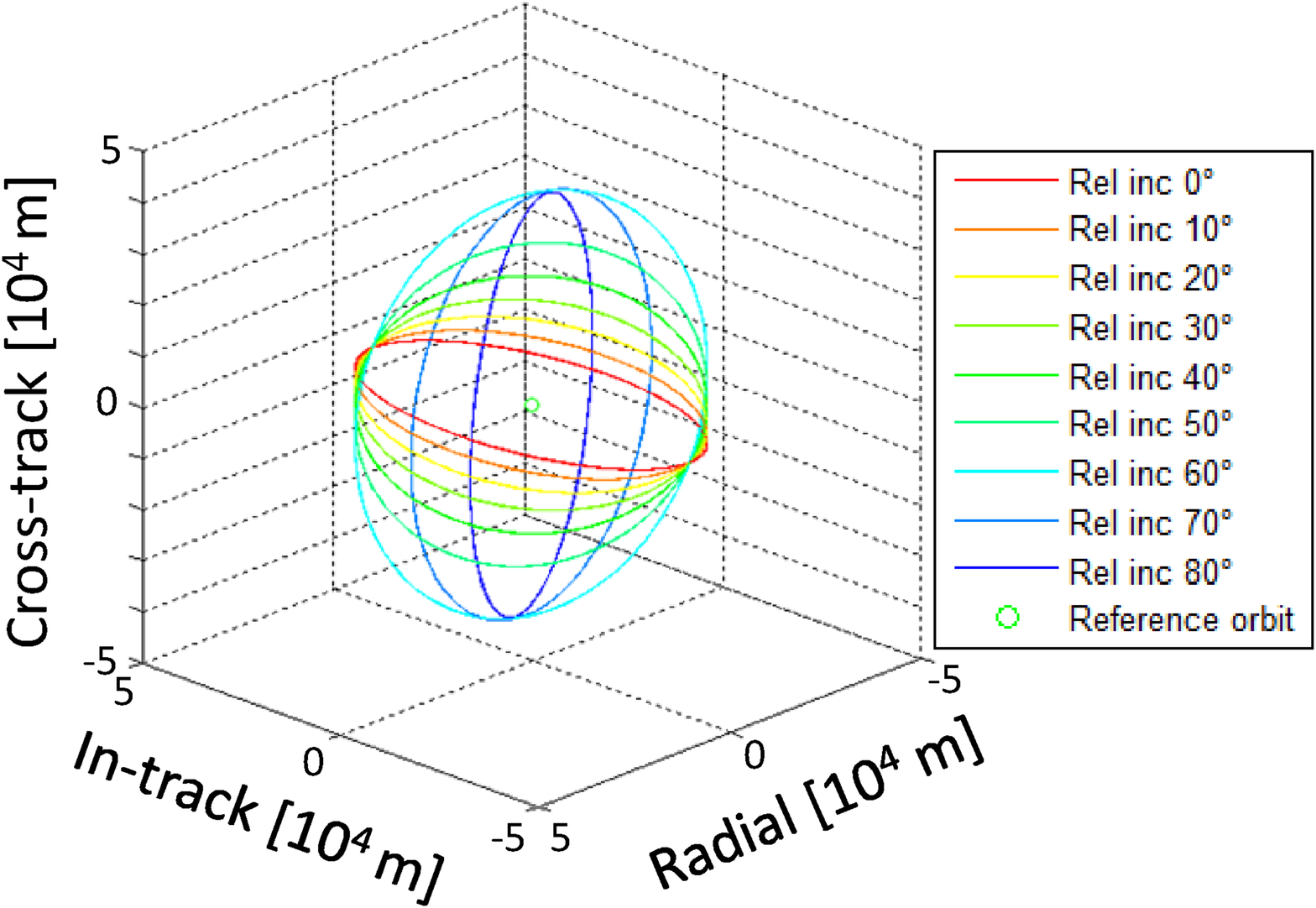}} \hspace{15mm}
\parbox{2in}{\includegraphics[scale=0.10]{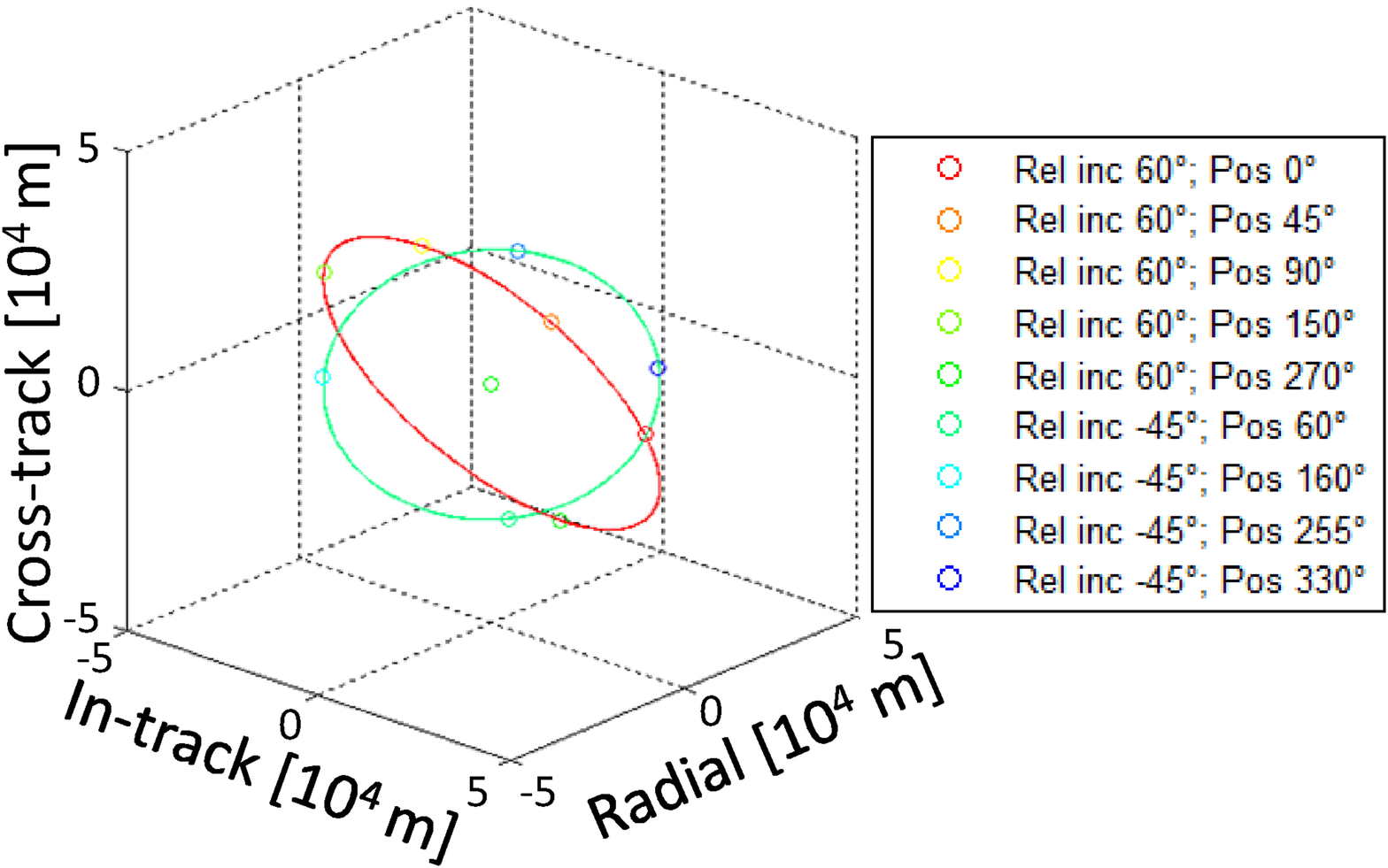}} 
\caption{\emph{\textbf{Orbiting the Moon:}} \footnotesize (Left) Relative orbits with different relative inclinations. (Right) Two relative orbits with several spacecraft in each orbit. These figures represent the relative orbits with respect to a reference orbit, with the Moon as the central body, without any perturbations.}
\label{fig:lunarOrbits}
\end{figure*}

% Moon far-side
% Hence deployment at this location could possibly allow for continuous observations, provided satellites can transmit their data through relay elements in Lunar orbit.

%The Earth-Moon L4 and L5 are much closer for Earth based communication, however are suspected to be radio quiet relative to the far side of the Moon. Alternatively, an array on the Moon far side would not only provide perfect shielding from made made RFI, but also high temperature and antenna gain stability \citep{wolt2012}. The Earth-Moon L2 located at \ssim$61347$km away from the Moon, is still in the cone of radio-silence and is sufficiently shielded from RFI. However this may not be a favorable deployment location since transmission in this radio quiet zone may affect future missions \citep{maccone2005}.

\subsection{Lagrange points and Moon-farside} The relative velocities of the satellites are minimal at Lagrangian points and hence these locations offer increased positional stability for longer time intervals. Therefore, the Lagrange points are an optimal choice to increase the integration time of the observations and also the mission lifetime. The Earth-Moon L4 and L5 are much closer for Earth based communication, however are suspected to be radio quiet relative to the Earth-Moon L2. The Earth-Moon L2 located at \ssim$61347$km away from the Moon, is still in the cone of radio-silence and is sufficiently shielded from RFI. However, this lagrange point may not be a favorable deployment location since transmission in this radio quiet zone may affect future missions \citep{maccone2005}. The Sun-Earth L4 and L5 points are too far and subsequently limit downlink rates. In contrast, the Sun-Earth L2 liberation point at \ssim$1.5$ Million km away from Earth,  is a tradeoff between downlink data rate, RFI avoidance and increasing $\tau_{int}$. Although this is a stationary point, in practice a satellite operating at L2 will experience a gravity gradient with a slow and steady outward drift. Such a scenario is preferred by the FIRST \citep{bergman2009} and SURO-LC \citep{baan2012} studies. The SURO-LC proposes a array of $8$ daughter satellites drifting slowly in Lissajous orbit and a mothership at a fixed distance of $10$km from the cluster. While such a mission will provide enhanced imaging performance with improved uvw coverage and longer integration times, the downlink data rate is estimated to be $2-3$ orders of magnitude less than an Moon based array using current technology \citep{rajan2011ac}.

\subsection{Orbiting the Moon} An equatorial orbit around the Moon presents a relatively easier down-link to Earth and sufficiently long eclipse times behind the Moon \wrt Earth. The long eclipse time periods shield against radio noise from Earth and enable the science observations. In the DARIS  study, to increase the predictability of the relative positioning, the reference orbit around the Moon was chosen to be circular which additionally also decreases the chance of collisions \citep{saks2010}. The array formation is build up from different relative orbits with a different relative inclination around a reference orbit (\figurename\ \ref{fig:lunarOrbits}(a)), where each of these relative orbits contain several node spacecrafts (\figurename\ \ref{fig:lunarOrbits}(b)). The reference orbit determines the duration of the eclipse time and subsequently the science duty cycle, which is increased by aligning an orbit with the Earth-Moon plane and/or by lowering the altitude \citep{boonstra2011daris}. As seen in \figurename\ \ref{fig:lunarOrbits}(c), the Eclipse time period can be increased by decreasing the orbital altitude, however consequently the percentage of the orbit in the shade increases. In addition, by decreasing the orbital altitude, the relative range rates of the satellites also increase, which in turn affects the baseline stability. Hence, a balance between the relative velocity and the eclipse time must be found. When including the perturbations of the Earths' gravity field, the irregularities in the lunar gravity field, the solar gravity field and the solar pressure, a constant drift of the relative orbits occurs. Coincidentally, this drift is mainly along the in-track direction of the reference orbit which can be compensated by adjusting the semi-major axis of the spacecraft node. In essence, a circular orbit in the Lunar equatorial plane offers a stable orbit, provided a trade-off is achieved between eclipse time and the satellite range rates.

\begin{figure*}[!t] 
\centering %\sidecaption
\psfrag{sE}[cc]{\small \textbf{Eclipse time (\%)}}
\psfrag{sA}[cc]{\small \textbf{Altitude (km)}}
\psfrag{sO}[tc]{\small \textbf{Orbital period (hours)}}
%\parbox{1.5in}{
%\includegraphics[scale=0.22]{figs/orbits/eclipse.eps} } 
\hspace{5mm}
\caption{\emph{\textbf{Altitude vs Orbital period in Lunar orbit:} } \footnotesize A satellite cluster orbiting the moon over the far-side enters a ``cone of silence'' behind the moon once every orbit. During this phase, the moon shields the satellite cluster from Earth-based and solar interference, thus permitting relatively noise-free scientific observations. These eclipse periods can be extended by inserting the satellite network at a higher altitude, which also offers more orbital stability \citep{rajan2011ac}. However, a longer eclipse period behind the moon and subsequently larger science duty-cycle constrains the communications with Earth.} \label{fig:lunarOrbits}
\end{figure*}

%\begin{figure*}[!t]
%\centering
%\includegraphics[scale=0.23]{figs/orbits/solar.eps}
%\caption{\emph{\textbf{Orbiting the Sun:}} \footnotesize (a) Reference orbit of the satellites with respect to the Earth (b) Relative orbits of the satellites with the mothership in the center over a three years period (c) Range rate for the satellite cluster in $60^\circ$ relative inclination plane \wrt a satellite cluster in the $-60^{\circ}$ plane at the $90^{\circ}$ position.}
%\label{fig:solarOrbits}
%\end{figure*}

\subsection{Orbiting the Sun} A potential reference orbit for formation flying around the Sun is the Earth orbit itself. However, if the satellites are too close to Earth, then the terrestrial interference is a major disturbance to science observations. Alternatively, an orbit around the Sun with a different eccentricity than the Earth orbit keeps the satellite array at $4$ to $10$ million km from Earth, which is far enough to offer both stability and also reduce radio noise from Earth. 
The large distance separation severely limits the available down-link bandwidth upto at least an order magnitude compared to the Lunar orbits. In view of an optimal balance between increased data-downlink and RFI free science observation, we choose the Earth orbit as a reference orbit with the satellite nodes orbiting at a distance of $4$ to $10$ million km from Earth. Hence, even though the constellation orbits the Sun as a central body, the reference orbit does go around the Earth, from leading to trailing. One of the many benefits of this particular orbit is that it is relatively stable for $10$ years and allows continuous scientific observations. However unlike the lunar orbital design, this reference orbit is eccentric and highly sensitive to small changes due to large difference between the semi-major axis and the relative pairwise distances of the satellites \citep{boonstra2011daris} \citep{saks2010}. Furthermore, the time period of the reference orbit is equal to the period of the relative orbits, which causes the formation to drift in all directions. With reference to the reference orbit, the cluster will slowly expand with time and hence offers unique sampling points for interferometry. One of the key advantages of this orbit is the low relative range rates which facilitates longer integration period. In addition, the relative range rates of the satellites in this orbit are less than $20$cm/s for $3$ years. Despite this advantage, the solar orbit is sensitive to small errors in velocity and the relative orbits are stable only for change in injection velocities upto $0.1$mm/s, which can be compensated using minor corrections \citep{saks2010}.

\section{Summary and Discussion} \label{sec:challenges} A satellite cluster of less than $10$ nodes is scientifically very interesting and meets the requirements for the extra-galactic survey science cases in terms of resolution and sensitivity. At least $4$ antennas observing at $30$ MHz for more than a year is sufficient to achieve the confusion limit of $65$ mJy with $~1'$ resolution, in which case over a million sources can be detected (Section \ref{sec:systemDefinition}). Moreover, even with fewer antennas, transient science cases such as bright Jupiter-like flares and Crab-like pulses can be addressed. All the satellites will be equipped with $2$ (or $3$) $5$m dipole antennas (or two $2.5$ monopoles) to observe the $\le 30$MHz spectrum (Section \ref{sec:antennas}).

For a nominal observational bandwidth of $\ge 1$MHz, each satellite is estimated to generate $\ge 6$ Mbits/s, which must be correlated in space to minimize downlink data rate to Earth. In both centralized and distributed scenarios, The processing requirements for filtering and correlation is negligibly small for upto $~50$ satellites and can be readily incorporated into the On Board Computer (OBC) (Section \ref{sec:processing}). To establish the Inter-satellite link, satellites will be equipped with patch antennas to transmit the desired $\ge 6$Mbps data rate. The ISL budget analysis shows that in the Centralized scenario using $2.45$GHz ISM band, the Node to mothership link can be established with $5$W over $50$km distance with a positive link margin. However, in the Distributed scenario upto $15$W is desired to establish a link over $100$km, which could be improved using clustering schemes and multi-hop communication (Section \ref{sec:communication}). Nonetheless, the proposed Distributed framework remains indispensable for large and scalable array of $\ge 10$ satellites, where SPOF must be avoided.

The on-board clock on all satellites must have an Allan-deviation of $\le 10^{-12}$, which can be met by a Rubidium or OCXO clock (Section \ref{sec:clocks}). The current bulky space-qualified clocks, such as EADS OCXO, will potentially be replaced by light-weight and low-power on-chip atomic clocks \eg SA.45s. In inaccessible (\eg Moon-farside) or far-away deployment scenarios (\eg Lagrange points), the satellites can be synchronized and localized using MGLS like algorithms, which enable the satellite network to be self-reliant co-operative network with minimum dependence on Earth-based ground stations (Section \ref{sec:syncLoc}). In addition, the orientation of the satellites can be estimated using the sensors in the Attitude and Orbit Control System (AOCS) which include the sun sensor and star trackers. All satellites will also be equipped with sufficient propulsion to ensure precise deployment and to maintain the maximum baseline separation of $100$km (Section \ref{sec:deploymentLocations}).

\begin{figure}[t!]
\centering

\includegraphics[scale=0.25]{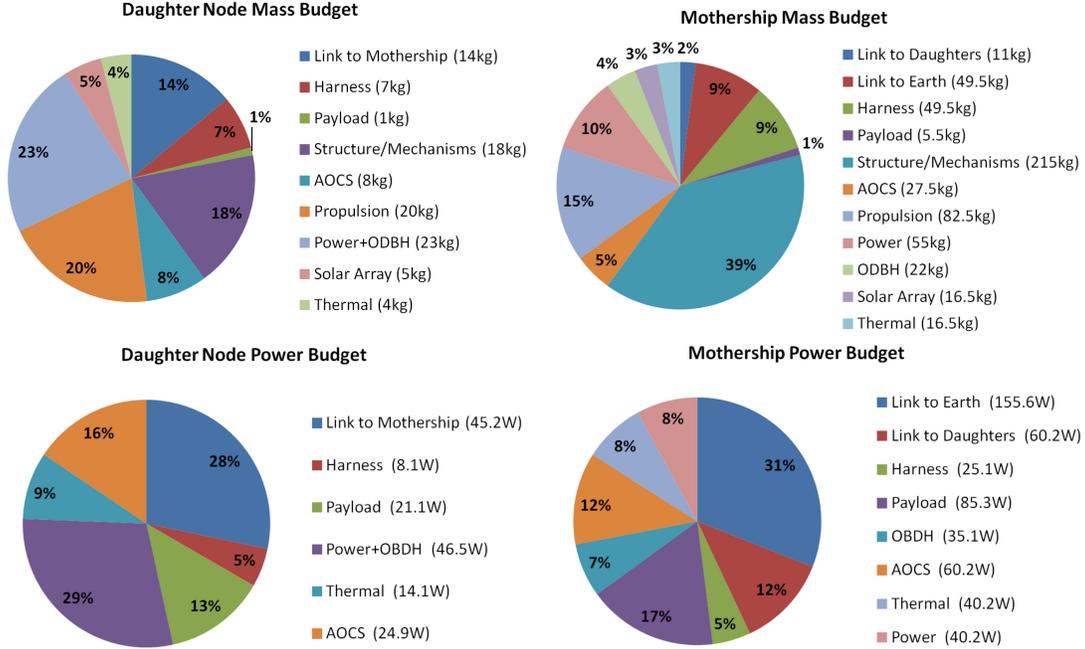}
\caption{\textbf{\emph{Mass and Power budget analysis of the DARIS mission:}} \footnotesize The DARIS mission consists of $8$ Daughter nodes and $1$ centralized mothership. The mass (and power) of each Daughter satellite and mothership was estimated to be $100$kg ($160$W) and $550$kg ($502$W) respectively.} \label{fig:satelliteMassPower} \end{figure}

\subsection{Technological challenges for ULW arrays} The actual satellite implementation is intricately connected to the specific mission requirements, the number of satellites, the active choices in network architecture and the deployment location. However, recent studies which investigated centralized scenarios for an ULW array give insights into the current state-of-the-art space technology. \figurename \ref{fig:satelliteMassPower} shows the mass and power breakdown for the DARIS mission, where all subsystems use only existing and tested off the shelf components \citep{boonstra2011daris}. The power consumption for the daughter node and the mothership was estimated at $160$W and $502$W respectively. Reliable and highly efficient solar panels based on triple junction GaAs cells were employed on both the mothership and Daughter nodes to meet the power requirements. Furthermore, the dry mass of each daughter node was estimated at \ssim$ 100$kg and the mothership at \ssim$ 550$kg. 

In comparison to DARIS, futuristic missions such as OLFAR, are expected to be lighter by two orders of magnitude and consuming an order of magnitude lesser power (see \tablename\ \ref{tb:studies}). The reduced mass and power requirements will not only enable a larger array of antennas for radio astronomy, but can potentially enable the system to piggy-back on other missions, without the need for a dedicated launch vehicle. Thus, future missions will possibly consist of relatively cheaper nano-satellites with miniaturized and power-efficient subsystems. %\color{red} Cubesats: DelfiC3 \color{black}

The intra-satellite communication between the satellite nodes is a fundamental bottleneck, which limits the bandwidth of observation and possibly the achievable baseline for radio astronomy imaging (see Section \ref{sec:islLinkmargin}). In addition to limiting the feasibility of the science cases, the power consumption of existing technologies is also high. The DARIS project indicates $>25 \%$ of the power consumption for communication for both the daughter node and mothership respectively (\figurename \ref{fig:satelliteMassPower}). The satellite network to Earth communication limits the possible number of satellites in the cluster. Moreover, one of the limiting factors for the number of satellites is the downlink data rate of the satellite network, such as the HBA XDU for the centralized architecture (see Section \ref{sec:earthDownlink}). In case of distributed architecture, the satellite-swarm will employ diversity schemes to cooperatively downlink data to earth \citep{budianu2014acta}.

Further potential research areas identified during the study include the antenna design for observation frequencies below $10$MHz, development of efficient imaging techniques for radio astronomy, high speed and robust RF Inter-satellite communications techniques \citep{budianu2013} and investigating control and reliability of large satellite arrays \citep{engelenTAES2014}. In addition, observability challenges such as the unknown RFI environment at the desired deployment location must also be investigated, possibly by a $\ge 2$ satellite interferometer via a precursor mission.

\subsection{Conclusion} The frequency window of $\le 30$ MHz opens a new realm of interesting science cases and yet remains the last unexplored frequency regime in astronomy. To achieve the science objectives at these wavelengths with desired resolution and sensitivity, a dedicated space-based ULW array is necessary. Recent advances in technology and computing resources have improved both the feasibility and scientific desirability of such a space-based array. In this article, we justified the need for a space-based antenna array for ultra-long wavelength radio astronomy and discussed various subsystems needed to achieve the desired science cases. More recently concluded projects such as DARIS, FIRST, SURO have shown feasibility of such an array. In particular, the DARIS project showed that a cluster of less than $10$ satellites can be launched using current off the shelf technology. An expanded set of science cases can be targeted by scaling the number of satellite nodes, extending the frequency range of observation and increasing the instantaneous bandwidth. However, this would significantly increase the mass, power consumption and eventually the cost of the mission. The on-going work on miniaturized nano-satellites may overcome this bottleneck and pave the way for feasible and affordable missions in the future.

\section{Acknowledgments} The authors would like to thank all the project members of DARIS, OLFAR and SURO for discussions, in particular Willem Baan, Heino Falcke and Kees van't Klooster. This research was funded in part by two projects namely, the STW-OLFAR (Contract Number: 10556) within the ASSYS perspectief program and secondly the ESA-DARIS Contract ``Feasibility of Very Large Effective Receiving Antenna Aperture in Space" (Contract Number: 22108/08/NL/ST).

%\newpage
\bibliographystyle{apalike} %natbib
\bibliography{myRef}
\end{document}